\documentclass[usenatbib]{mn2e}

\usepackage{times}
\usepackage{amssymb}
\usepackage{rotating}
\usepackage{xspace}
\usepackage{caption}
\usepackage{amsmath}
\input{epsf}

\title[Reaching the Peak of the Quasar SED II]
{Reaching the peak of the quasar spectral energy distribution -- II. Exploring the accretion disc, dusty torus and host galaxy}

\author[James\,S. Collinson et al.]{James S. Collinson$^1$\thanks{Email: j.s.collinson@durham.ac.uk}, Martin J. Ward$^1$, Hermine Landt$^1$, Chris Done$^{1}$, Martin Elvis$^2$ \newauthor and Jonathan C. McDowell$^2$\\
 \\
$^1$Centre for Extragalactic Astronomy, Department of Physics, University of Durham, South Road, Durham, DH1 3LE, UK\\
$^2$Harvard-Smithsonian Center for Astrophysics, 60 Garden St., Cambridge, MA 02138, USA\\
}

\pagerange{\pageref{firstpage}--\pageref{lastpage}} \pubyear{2014}

\newcommand{\xmm}{{\it XMM}\xspace}
\newcommand{\xmmn}{{\it XMM-Newton}\xspace}

\newcommand{\hst}{{\it HST}\xspace}

\newcommand{\galex}{{\it GALEX}\xspace}

\newcommand{\optxagnf}{{\sc optxagnf}\xspace}
\newcommand{\optxconv}{{\sc optxconv}\xspace}
\newcommand{\wabs}{{\sc wabs}\xspace}

\newcommand{\zwabs}{{\sc zwabs}\xspace}

\newcommand{\zdust}{{\sc zdust}\xspace}
\newcommand{\xspec}{{\sc xspec}\xspace}

\newcommand{\python}{{\sc python}\xspace}

\def\Lbol{$L_{\rm bol}$\xspace}
\def\LEdd{$L_{\rm Edd}$\xspace}
\def\Msun{$M_{\odot}$\xspace}
\def\risco{$r_{\rm isco}$\xspace}
\def\rout{$r_{\rm out}$\xspace}
\def\rsg{$r_{\rm sg}$\xspace}
\def\rcor{$r_{\rm cor}$\xspace}
\def\Rg{$R_{\rm g}$\xspace}

\def\astar{$a_*$\xspace}

\def\mdot{{$\dot{m}$}\xspace}
\def\M_BH{$M_{\rm BH}$\xspace}

\def\rchi{{$\chi^{2}_{\rm red}$\xspace}}

\def\H0{{\rm ~km~s^{-1}~Mpc^{-1}}}

\def\eg{{e.g.\ }}

\def\ie{{i.e.\ }}

\def\la{\mathrel{\hbox{\rlap{\hbox{\lower4pt\hbox{$\sim$}}}{\raise2pt\hbox{$<$}}
}}}
\def\ga{\mathrel{\hbox{\rlap{\hbox{\lower4pt\hbox{$\sim$}}}{\raise2pt\hbox{$>$}}
}}}
\def\d25{$D_{25}$}
\def\nh{{$N_{\rm H}$}}
\def\Lya{{Ly-$\alpha$}\xspace}
\def\Hi{{H$\,${\sc i}}\xspace}
\def\Ha{{H$\alpha$}\xspace}
\def\Hb{{H$\beta$}\xspace}
\def\Hg{{H$\gamma$}\xspace}
\def\Hd{{H$\delta$}\xspace}
\def\Hei{{He$\,${\sc i}}\xspace}
\def\Heii{{He$\,${\sc ii}}\xspace}
\def\Cii{{C$\,${\sc ii}]}\xspace}
\def\Ciii{{C$\,${\sc iii}]}\xspace}
\def\Civ{{C$\,${\sc iv}}\xspace}
\def\Mgii{{Mg$\,${\sc ii}}\xspace}
\def\Oi{{O$\,${\sc i}}\xspace}
\def\Oii{{[O$\,${\sc ii}]}\xspace}
\def\Oiii{{[O$\,${\sc iii}]}\xspace}
\def\Oiv{{O$\,${\sc iv}]}\xspace}
\def\Neiii{{[Ne$\,${\sc iii}]}\xspace}
\def\Neiv{{[Ne$\,${\sc iv}]}\xspace}
\def\Aliii{{Al$\,${\sc iii}}\xspace}
\def\Feii{{Fe$\,${\sc ii}}\xspace}

\def\Siii{{Si$\,${\sc ii}}\xspace}
\def\Siiv{{Si$\,${\sc iv}}\xspace}

\def\hi{H{\sc i}\xspace}

\begin{document}

\maketitle

\label{firstpage}

\begin{abstract}

We continue our study of the spectral energy distributions (SEDs) of 11 AGN at \mbox{$1.5<z<2.2$}, with optical-NIR spectra, X-ray data and mid-IR photometry. In a previous paper we presented the observations and models; in this paper we explore the parameter space of these models. We first quantify uncertainties on the black hole masses (\M_BH) and degeneracies between SED parameters. The effect of BH spin is tested, and we find that while low to moderate spin values ($a_* \leqslant 0.9$) are compatible with the data in all cases, maximal spin ($a_* = 0.998$) can only describe the data if the accretion disc is face-on. The outer accretion disc radii are well constrained in 8/11 objects, and are found to be a factor $\sim 5$ smaller than the self-gravity radii. We then extend our modelling campaign into the mid-IR regime with \textit{WISE} photometry, adding components for the host galaxy and dusty torus. Our estimates of the host galaxy luminosities are consistent with the \M_BH--bulge relationship, and the measured torus properties (covering factor and temperatures) are in agreement with earlier work, suggesting a predominantly silicate-based grain composition. Finally, we deconvolve the optical-NIR spectra using our SED continuum model. We claim that this is a more physically motivated approach than using empirical descriptions of the continuum such as broken power-laws. For our small sample, we verify previously noted correlations between emission linewidths and luminosities commonly used for single-epoch \M_BH estimates, and observe a statistically significant anti-correlation between \Oiii equivalent width and AGN luminosity.

\end{abstract}

\begin{keywords}
black hole physics; accretion discs; quasars: supermassive black holes, emission lines; galaxies: active, high-redshift
\end{keywords}

\section{Introduction} \label{sec:intro}
In an active galactic nucleus (AGN), accretion of gas onto a central galactic supermassive black hole (BH) releases a large amount of energy across a broad wavelength range. The broad-band spectral energy distribution (SED) of this luminous accretion flow is shaped by the BH properties, as well as the structure and orientation of the infalling matter. Interpreting the observed properties of AGN SEDs as the result of known physical phenomena enables us to address key questions about these energetic objects. These include how the BH grows over cosmic time, the poorly understood mechanism by which relativistic jets are formed and driven, and the role AGN play in their host galaxies, in particular with respect to feedback via winds and outflows, which are thought to contribute significantly to galaxy formation (\eg \citealt{king10}, \citealt{mccarthy10}, \citealt{nardini15}). Much effort has therefore been directed at researching AGN through studies of their SEDs, both by acquiring better quality data, and developing increasingly advanced SED models (\eg \citealt{ward87}, \citealt{elvis94}, \citealt{vasudevan09}, \citealt{jin12_1}, \citealt{netzer14}).

\begin{figure*}
	\centering
	\includegraphics[trim=0cm 0cm 0cm 0cm, clip=true, width = 0.9\textwidth]{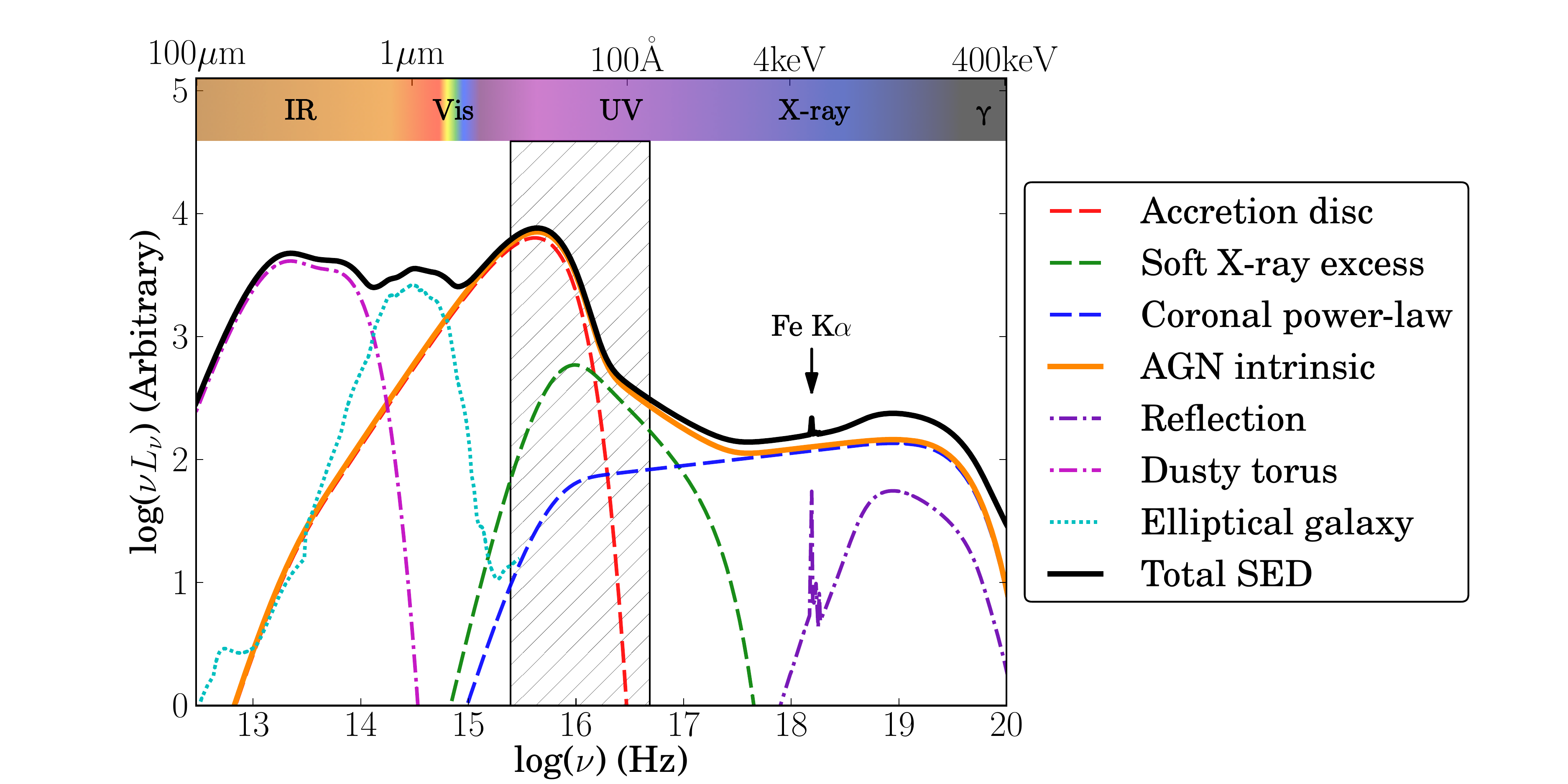}
	\caption{\small A simplified schematic diagram of an AGN SED, showing the approximate shape and extent of the various components discussed in the text. Dashed lines denote AGN intrinsic emission, dash-dot lines show emission reprocessed by the surrounding material and the dotted line shows starlight from an elliptical host galaxy. The hatched region highlights the spectral range that is heavily obscured by absorption in the IGM. We only show an elliptical galaxy here; galaxies with active star formation have stronger UV/IR contributions.}
	\label{fig:sed}
\end{figure*}

In the optical-UV, the SED comprises a `big blue bump,' which is broadly consistent with that expected from an optically thick, geometrically thin \cite{shakura73} accretion disc (AD; see also \citealt{novikov73}, \citealt{czerny87}). At infrared (IR) wavelengths, a significant contribution from a hot, dusty torus is observed (\eg \citealt{barvainis87}, \citealt{pier93}, \citealt{mor09}). This structure offers an explanation for the dichotomy between Type 1 and Type 2 AGN in the \cite{antonucci93} unified model, as the torus obscures the central engine from view in the latter class. The X-ray energy range is dominated by a power-law tail (PLT), thought to arise from inverse Compton scattering by a hot, optically thin corona (\eg \citealt{haardt91}), with additional contributions by a hard X-ray reflection hump (\citealt{george91}, \citealt{done10}), and the so-called soft X-ray excess (SX). The origin of the SX is uncertain; it may be caused by further X-ray reflection off the AD (\citealt{crummy06}), partially ionised absorption (\citealt{gierlinski04}), or warm Compton upscattering within the AD (\citealt{done12}). A schematic of the different components that make up the AGN SED is shown in Fig.\ \ref{fig:sed}.

A number of properties of AGN influence the observed SED. The matter that accretes onto the BH is net-neutral, so the BH has just two intrinsic properties -- the mass (\M_BH) and spin (\astar) -- which primarily affect the peak temperature and mass-energy conversion efficiency in the AD. The rate of mass accretion through the AD (usually given in terms of the Eddington ratio, \mbox{$\dot{m}=L_{\rm bol}/L_{\rm Edd}$}, where \Lbol is the AGN bolometric luminosity and $L_{\rm Edd}$ is the Eddington-limited luminosity) and its orientation with respect to the observer ($\theta$) can further modify the SED, and photoelectric absorption and dust extinction along the line of sight attenuate the emission over large wavelength ranges.

\M_BH can be estimated via a number of methods in AGN. Reverberation mapping (RM) uses the time delay between changes in the ionising continuum and the broad line region (BLR) to probe the central potential, and estimate \M_BH (\eg \citealt{blandford82}, \citealt{peterson93}, \citealt{kaspi00}, \citealt{peterson04}, \citealt{bentz13}, \citealt{shen16}), but this is an observationally intensive exercise. RM has allowed the calibration of the `single-epoch virial mass estimation' method, where broad line profiles and luminosity measurements from single spectra yield a (less accurate) estimate of \M_BH (\eg \citealt{wandel99}, \citealt{kaspi05}, \citealt{vestergaard02}, \citealt{matsuoka13}). The choice of emission line and continuum measurement has been shown to be important, as some line profiles are shaped by effects such as outflows that make them susceptible to bias \citep{shen12}. It has also been suggested that $\dot{m}$ can be estimated from spectral continuum measurements via so-called bolometric correction (BC) coefficients (\eg \citealt{mclure04}, \citealt{vasudevan07}, \citealt{trakhtenbrot12}).

BH spin affects the radius of innermost stable circular orbit (\risco) around the BH. For $a_*=0$ (non-spinning) BHs, \mbox{$r_{\rm isco} = 6 R_{\rm g}$} (where $R_{\rm g} = GM_{\rm BH}/c^2$ is the gravitational radius), but this decreases with increasing spin, so at $a_*=0.998$ (maximal spin), $r_{\rm isco} \simeq 1.24 R_{\rm g}$. Previous work has constrained \risco, and hence \astar, by fitting the profile of the broad, relativistically distorted, Fe K$\alpha$ emission line that is often observed in the X-ray reflection spectrum of AGN (see Fig.\ \ref{fig:sed}). This method requires high signal-to-noise (S/N) X-ray data, limiting its application to nearby, bright AGN (\citealt{fabian89}, \citealt{fabian09}, \citealt{risaliti13}, \citealt{reynolds14}). Moreover, this technique is contentious, as it requires an extreme X-ray source geometry to sufficiently illuminate the inner part of the AD (\eg \citealt{zoghbi10}). Alternative models propose that the profile of the K$\alpha$ line could be influenced by complex, multi-layered absorption, possibly from a disc wind (\eg \citealt{miller07}, \citealt{turner09}, \citealt{miller13}, \citealt{gardner14}).

Due to these uncertainties, attempts have been made to constrain the BH spin by fitting theoretical AD models to data. These models are generally based on the description by \cite{shakura73}, with general relativistic, colour temperature and radiative transfer corrections applied (\citealt{hubeny00}, \citealt{davis06}). Various modifications can be added, such as winds and outflows \citep{slone12}. These should be associated especially with high accretion rates, where photon trapping within the disc can also become important (leading to a `slim disc,' \eg \citealt{abramowicz88}, \citealt{wang14}). However these models are not well understood (\citealt{castello-mor16}), and are beyond the scope of this paper.

\cite{capellupo15} and \cite{capellupo16} used a thin AD model to constrain spin values in their sample of 39 AGN at $z \sim 1.5$. Using the numerical code of \cite{slone12}, and contemporaneous optical--near-infrared (NIR) spectra from the VLT/X-shooter instrument, they successfully modelled the rest-frame optical-UV SED in all but two objects. The most massive BHs in their sample have the highest measured spin values, supporting a `spin-up' description of AGN BH evolution, where prolonged unidirectional accretion episodes and BH mergers increase the spin of BHs through cosmic time (\eg \citealt{dotti13}).

\cite{done13} and \cite{done15} explored the spins of two low-mass, local AGN -- PG1244$+$026 and 1H 0707$-$495 respectively -- by applying the Done et al.\ (2012, 2013) SED model to multiwavelength data. They found that both objects were highly super-Eddington if modelled with high spin values, implying that the underlying AD model assumptions break down. They argue that the disc is unlikely to radiate all the liberated gravitational energy, due to winds and/or advection, meaning that its peak temperature and luminosity no longer give robust constraints on spin. Additionally, this means that it is unlikely the disc is flat, which is the geometry assumed in the derivation of BH spin from the Fe K$\alpha$ line \citep{fabian09}.

\cite{wu13} and \cite{trakhtenbrot14} also used SED-based arguments to probe the spins and assembly histories of BHs in AGN. Both works specifically focussed on inferring the accretion efficiency, $\eta$, which increases with the BH spin. While \cite{wu13} found no significant correlation between \M_BH and radiative efficiency (and hence spin), \cite{trakhtenbrot14} did find such a relation, with most of the extremely massive AGN in their sample having efficiencies corresponding to high spins. However, \cite{raimundo12} found that it is extremely difficult to accurately determine the efficiency via such means.

The nature of the putative dusty torus is the subject of considerable debate. Studies of the extent, composition and dynamics of this structure make use of spectrophotometric observations, and time dependent variability. Previous work has found evidence that the torus could be `clumpy' in nature (\eg \citealt{nenkova02}, \citealt{dullemond05}, \citealt{nenkova08}), but this is as yet uncertain (\eg \citealt{lawrence10}). Of particular interest are the peak dust temperature in the torus, which can be used to infer its composition (\citealt{netzer15}), and the total luminosity, from which the dust covering factor can be estimated.

In local AGN, a remarkable relationship between the host galaxy and the central BH has been observed. These include strong correlations between \M_BH and the stellar velocity dispersion of the galaxy (\M_BH--$\sigma$; \eg \citealt{ferrarese00}, \citealt{gebhardt00}, \citealt{beifiori12}) and between \M_BH and the bulge mass/luminosity (\eg \citealt{magorrian98}, \citealt{marconi03}, \citealt{sani11}). Whilst local galaxies can be easily resolved in imaging, enabling structural decomposition of the point-like AGN and extended galaxy bulge and disc (\eg \citealt{marconi03}, \citealt{mcconnell13}), disentangling these contributions is challenging at higher redshifts. \cite{peng06} partially overcame this limitation by using \textit{Hubble Space Telescope} (\hst) imaging of gravitationally lensed AGN, and found evidence that at $z>1.7$, the \M_BH--bulge mass ratio is $\gtrsim4$ times that observed locally.

In our previous paper (\citealt{collinson15}, hereafter Paper I) we presented a means of systematically modelling the SED of a sample of 11 medium redshift ($1.5 < z < 2.2$) AGN, using multiwavelength spectral data from IR to X-ray wavelengths and a numerical SED code described in \cite{done12}. In this sample, the redshift effect, and selection bias toward more massive AGN (\eg \citealt{mclure04}) that contain cooler accretion discs allowed us to sample the peak of the SED in five objects. This allowed us to make robust estimates of the AGN bolometric luminosity (\Lbol), noting that in several objects the SX was unconstrained by the available data. We also found that the host galaxy starlight contribution to the SED peak was small, but may be non-negligible at longer wavelengths, in the rest-frame optical spectrum. It was necessary to model host galaxy attenuation in the form of dust reddening and photoelectric absorption (\citealt{netzer79}, \citealt{jin12_1}, \citealt{castello-mor16}).

In this paper, we will extend and examine the parameter space of our models. In each object, we consider the contributions of six emissive components to the total SED; three components for the AGN central engine itself, the host galaxy and dusty torus, and the broad line region (BLR). We also consider attenuation by dust and gas in the Milky Way (MW) and host galaxy. We will:

\begin{itemize}
\item[i.]{Test the model properties, including both extrinsic effects (host galaxy extinction curves) and intrinsic effects, \eg spin and the uncertainties on \M_BH (Section \ref{sec:sedmodel}).}
\item[ii.]{Carry out an analysis of the toroidal dust component, using mid- and far-IR photometry from \textit{WISE} (Section \ref{sec:torus}).}
\item[iii.]{Consider the effects of variability in the optical spectra (Section \ref{sec:variability}).}
\item[iv.]{Undertake an optical/IR spectral decomposition, using our refined models of the underlying continua, and compare results from this approach to earlier studies (Section \ref{sec:spec_decomp}).}

We assume a flat cosmology with $H_0 = 70$ km s$^{-1}$ Mpc$^{-1}$, $\Omega_{\rm M} = 0.27$ and $\Omega_{\Lambda} = 0.73$ throughout.

\end{itemize}

\section{Testing the SED Model} \label{sec:sedmodel}

\subsection{Motivation} \label{subsec:motiv}

We discussed several limitations of our modelling campaign in Paper I, and begin this study by addressing these. In that work, we determined that for our data quality and coverage, using an SED model with intrinsic reddening/photoelectric absorption as free parameters was statistically justified. We also found that the data were generally good enough to determine the relative luminosities of the SX and PLT, though could not independently constrain the detailed shape of the SX. 

We modelled the intrinsic (\ie host galaxy) extinction using a redshifted \cite{cardelli89} MW extinction curve, with \mbox{$E$($B-V$)} as a free parameter. We concluded that not all of the AGN in our sample were well described by this model, a finding also noted by others. \cite{hopkins04} and \cite{glikman12} found the Small Magellanic Cloud (SMC) extinction curve to better describe host galaxy reddening in AGN, whilst \cite{capellupo15} found the SMC curve to be no better than that of the MW, or a simple power-law. \cite{castello-mor16} opted to use an SMC curve, but noted that the limited data coverage in the rest-UV made favouring any one model impossible. We therefore briefly explored using two alternative extinction curves -- SMC and Large Magellanic Cloud (LMC) -- and found that the intrinsic extinction in some objects was better characterised by these. In this work we will use this as a means of refining our existing models.

Throughout Paper I we kept \M_BH fixed at values calculated from the profile of \Ha, using the method of \cite{greene05}. However, it is known that such single epoch virial mass estimates can be uncertain, and we have yet to explore the effect of changing \M_BH on \Lbol, and other SED properties. This will be quantified in this paper. The estimated error on these \M_BH estimates, arising from the dispersion in the method, is $\sim \! 0.1$ dex, with a smaller additional contribution from measurement errors. In practice, the dominant sources of error on single epoch virial \M_BH estimates are the uncertainties in the BLR size--luminosity relationship and virial coefficient, estimated to contribute a $\sim 0.46$ dex total uncertainty \citep{park12}.

Similarly, we found that data in all objects could be modelled by keeping \astar fixed at zero (\ie non-spinning). This does not necessarily preclude higher spin values, so in this paper we will specifically test $a_* \neq 0$ scenarios.

Finally, in many objects, we observed that the outer AD radius (\rout) could be estimated from the SED fitting routine. However, we did not compare the values measured with the radius at which self-gravity within the AD becomes significant. Here, we will test two other means of handling \rout; firstly by fixing it to the self-gravity radius, and secondly by fixing it to an arbitrarily large value.

\subsection{Data and SED construction} \label{subsec:sed_construct}

In this paper we are primarily concerned with the nature of the underlying AGN SED continuum. In Paper I we described our sample selection, which focussed on the need for optical, NIR and X-ray spectra. To make a reliable \M_BH estimate, we required the broad Balmer emission lines, \Hb and \Ha, to lie in the NIR $J$ and $H$ bands. Our primary sample was therefore at $1.49 < z < 1.61$, and we included an additional object at $z \simeq 2.2$ which had the requisite data and \Hb and \Ha in the NIR $H$ and $K$ bands. The objects' names, positions, and other key properties are presented in Paper I, and we retabulate the names and mass estimates from \Ha in Section \ref{subsec:intred} of this work.

Our data come from four observatories. The optical spectra were extracted from the Sloan Digital Sky Survey (SDSS; \citealt{york00}) and Baryon Oscillation Spectroscopic Survey (BOSS) databases. NIR spectra for 7 objects were obtained using the Gemini Near-Infrared Spectrograph (GNIRS; \citealt{elias06}) and an additional 4 objects from the \cite{shen12} sample had pre-existing spectra from ARC's TripleSpec (TSpec; \citealt{wilson04}) instrument, kindly provided by Yue Shen. Finally, X-ray spectra were retrieved from the \xmmn (\citealt{jansen01}) science archive. We describe the data reduction in Paper I. The optical/IR spectra are corrected for MW extinction using the dust maps of \cite{schlegel98} and extinction law of \cite{cardelli89}.

We construct our SED using spectral data from all of these sources, following the same approach as in Paper I. The optical/IR spectra include a number of emission features, with the underlying continuum dominated by the various emission components shown in Fig.\ \ref{fig:sed}. For all SED fitting, we thus define continuum regions (free from emission line/Balmer continuum/blended \Feii emission) in the optical/IR spectra using the template of \cite{vandenberk01} as a guide. The optical--NIR continuum regions used are shown in Table \ref{tab:cont_wav}, with regions showing contamination by noise/emission features adjusted or removed accordingly. There is general consistency with other work that define similar such bandpasses, see \cite{kuhn01} and \cite{capellupo15}. Some of these wavelength ranges have been adjusted by a small amount from those used in Paper I, reflecting improved attempts to mitigate against inclusion of emission-contaminated ranges. In this section we do not include the continuum redward of \Ha (regions 10--15 in Table \ref{tab:cont_wav}) owing to the `red excess' seen in this region. This region is discussed and modelled later, see Section \ref{sec:torus}.

The X-ray spectrum is also known to show emission features, such as the previously mentioned Fe K$\alpha$ line, but the S/N of our X-ray data is not sufficient to resolve such features. We therefore use all available data from the MOS1, MOS2 and PN detectors (\citealt{turner01}, \citealt{struder01}) of the European Photon Imaging Camera (EPIC) aboard \xmmn to maximise the number of X-ray counts.

\begin{table}
	\caption{\small Optical/IR continuum regions used in the SED fitting. Not all regions were used in all objects, as some showed contamination by emission features such as \Feii. We avoided oversampling any part of the spectrum, aiming for a roughly even spread of continuum points across the spectral range (particularly with respect to regions 10--15). Before Section \ref{sec:torus}, only regions 1--9 are used for fitting, as regions 10--15 begin to show potential contribution by hot dust and host galaxy. Some of regions 10--15 are used in for modelling in Section \ref{sec:torus}.}
	\small
	\centering
	\begin{tabular}{cccc}
		\hline
		
		Region \# & Centre & Start & End \\
		& \multicolumn{3}{c}{(\AA, rest-frame)} \\
		\hline
		1 & 1325 & 1300 & 1350 \\
		2 & 1463 & 1450 & 1475 \\
		3 & 1775 & 1750 & 1800 \\
		4 & 2200 & 2175 & 2225 \\
		5 & 4025 & 4000 & 4050 \\
		6 & 4200 & 4150 & 4250 \\
		7 & 5475 & 5450 & 5500 \\
		8 & 5650 & 5600 & 5700 \\
		9 & 6100 & 6050 & 6150 \\
		10 & 7000 & 6950 & 7050 \\
		11 & 7250 & 7200 & 7300 \\
		12 & 7538 & 7475 & 7600 \\
		13 & 7850 & 7800 & 7900 \\
		14 & 8150 & 8100 & 8200 \\
		15 & 8900 & 8800 & 9000 \\
		
		\hline
		
	\end{tabular}
	\label{tab:cont_wav}
\end{table}

A limitation of our study is that the optical/IR/X-ray data were not collected contemporaneously. AGN are known to vary across all wavelengths differentially, and we therefore cannot rule out variability having occurred between observations. This is a limitation of most such studies due to the difficulty and expense of scheduling simultaneous observations using multiple space- and ground-based observatories. In Paper I we described our means of checking for evidence of variability between optical/IR observations using photometry and simple power-law continuum fits. We concluded that in one object (J0041$-$0947) there was evidence for $\sim30$ per cent variability between optical/IR observations. In this case, we used the GNIRS spectrum as observed for estimating \M_BH, but scaled it to the level of SDSS for SED fitting. In two objects with multiple epochs of X-ray data we found no evidence for X-ray variability, but cannot rule out a variable X-ray component in any of our AGN.

As discussed in Paper I, we do not use photometric data in our SED modelling. In the optical/IR, photometry is usually contaminated by emission features, and is inferior to the spectra for defining the continuum. \galex and \xmm optical monitor (OM) UV photometry is available for some objects (see Table E1 of Paper I), however, due to the redshift range of our AGN, these data lie on or beyond Ly-$\alpha$. Photometry covering Ly-$\alpha$ cannot be corrected, because the equivalent width of this strong feature varies widely between objects (\citealt{elvis12}). Similarly, photometry beyond Ly-$\alpha$ in the rest frame cannot be reliably used, as it is unpredictably attenuated by the multitude of narrow absorption features in the \Lya forest. Continuum recovery in this region requires high resolution UV spectra, \eg from \hst (\citealt{finn14}, \citealt{lusso15}).

Throughout this work we use the AGN SED model \optxagnf, described in \cite{done12}. This model comprises three components -- AD, SX and PLT -- and applies the constraint of energy conservation to these. We do not include a relativistic reflection component (see Fig.\ \ref{fig:sed}), as our \xmm spectra lack the S/N and coverage necessary to model this component. All SED fitting is performed in the \xspec spectral analysis package (\eg \citealt{arnaud96}), using a Levenberg-Marquardt minimisation routine.

\subsection{Model Refinement: Intrinsic Reddening} \label{subsec:intred}

In this section we refine our best models from Paper I by applying alternative intrinsic extinction curves. We will therefore produce a new SED model for each object, using newly determined best-fitting parameters.

There are several processes suppressing the observed flux of our data, so we must combine the \optxagnf model with attenuation components. We have already corrected the optical--NIR spectra for MW extinction (Section \ref{subsec:sed_construct}), but photoelectric absorption by neutral gas in the ISM absorbs the UV to soft X-ray part of the SED (see Fig.\ \ref{fig:sed}). In \xspec, we incorporate the multiplicative \wabs model (\citealt{morrison83}) to correct for this absorption, with \nh \ column densities taken from \cite{kalberla05}.

\begin{figure}
	\centering
	\includegraphics[trim=0cm 0.0cm 0cm 0.0cm, clip=true, width = 0.5\textwidth]{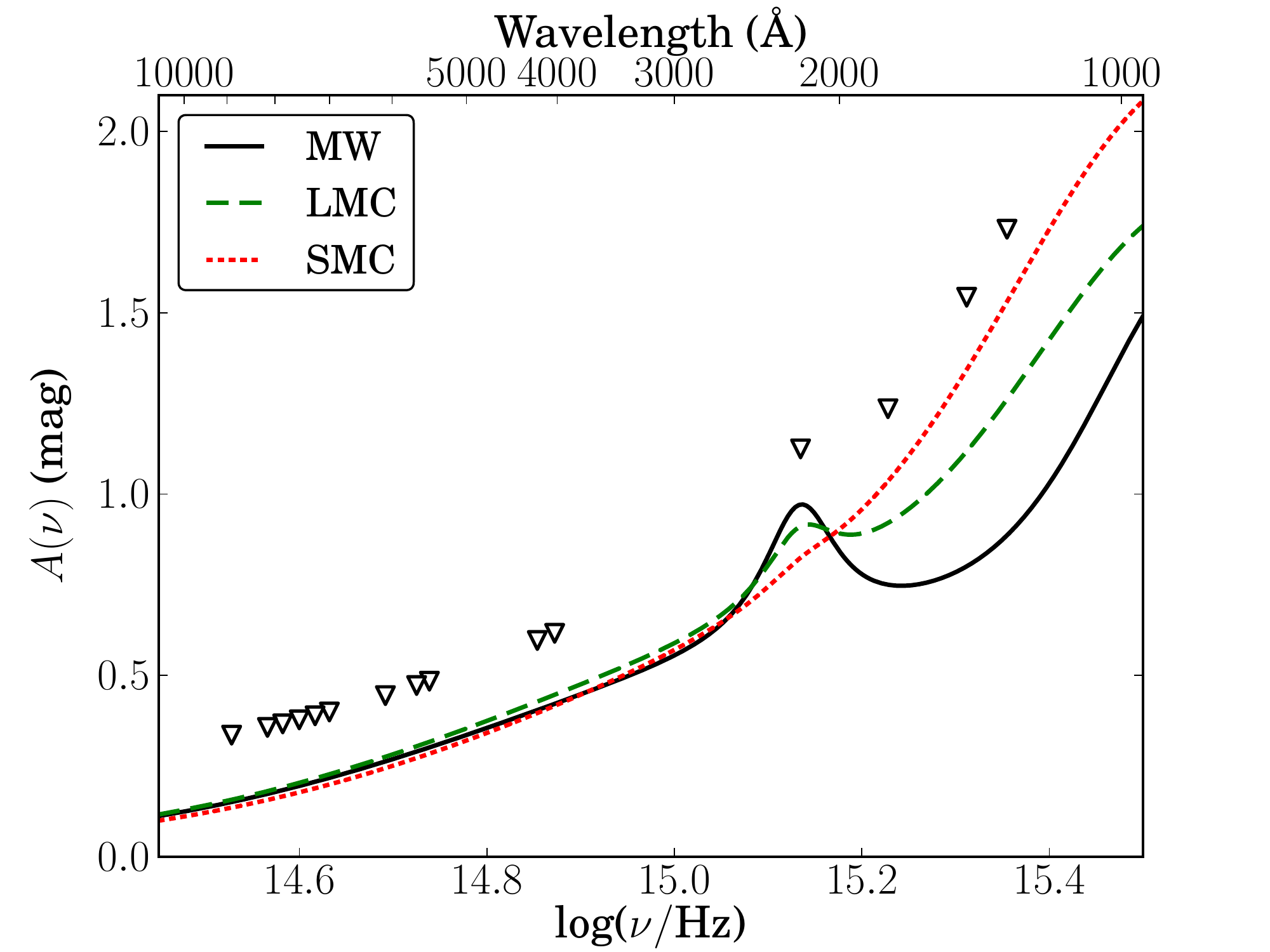}
	\caption{\small The three extinction curves we test for the intrinsic reddening. All curves are for the same $E$($B-V$). The markers indicate the locations of the continuum regions used for the fit. Clearly, the four bluest regions are the most important for distinguishing between the curves.}
	\label{fig:ext_curves}
\end{figure}

These same processes occur in the host galaxy, however, the extinction value and \nh \ column density cannot be measured directly. These parameters must therefore be left free in the SED fitting. We use \zwabs \ (a redshifted version of \wabs) for the host galaxy photoelectric absorption, and \zdust (a redshifted extinction curve) to model the host galaxy extinction. \zdust has a choice of three empirical reddening profiles -- MW, LMC and SMC -- which are described in \cite{pei92}, and shown in Fig.\ \ref{fig:ext_curves}. These models are similar at optical wavelengths, but show large differences in the UV range.

The full \xspec model therefore takes the form \mbox{\wabs $\times$ \zwabs $\times$ \zdust $\times$ \optxagnf}. A full table of the parameters of these models, together with starting values and limits, is presented in Appendix \ref{sec:app}. To summarise, the fixed properties are as follows:

\begin{itemize}
	\item[(i)] Mass, \M_BH: fixed at value from \Ha.
	\item[(ii)] Redshift, $z$.
	\item[(iii)] Comoving distance, $r_{\rm c}$.
	\item[(iv)] Spin, \astar: fixed at 0.
	\item[(v)] SX electron temperature, $kT_{\rm e}$: fixed at 0.2 keV.
	\item[(vi)] SX optical depth, $\tau$: fixed at 10.
	\item[(vii)] Ratio of absolute extinction to that defined by ($B-V$), \mbox{$R(V)=A(V) / E(B-V)$}: fixed at values of 3.08, 3.16, 2.93 for MW, LMC, SMC respectively (\citealt{pei92}).
\end{itemize}

The free parameters are:

\begin{itemize}
	\item[(i)] Mass accretion rate, \mbox{$\dot{m}=L_{\rm bol}/L_{\rm Edd}$}.
	\item[(ii)] Coronal radius, \rcor.
	\item[(iii)] Outer AD radius, \rout.
	\item[(iv)] PLT photon index, $\Gamma$.
	\item[(v)] Fraction of energy released below $r_{\rm cor}$ which powers SX rather than PLT, $f_{\rm SX}=(1-f_{\rm PLT})$.
	\item[(vi)] Intrinsic \hi column density, $N_{\rm H, int}$.
	\item[(vii)] Intrinsic reddening, $E$($B-V$)$_{\rm int}$.
\end{itemize}

We fit the model to all 11 objects for each of the MW, LMC and SMC extinction curves, and use the final \rchi \ fit statistic to gauge which produces the best fit to the data. We find that six objects are best described by the MW extinction curve, four by the LMC curve and one by the SMC curve. In objects for which the intrinsic reddening is low \mbox{($E(B\!-\!V)_{\rm int} \lesssim 0.03$ mag)}, the difference in \rchi \ is generally small, but these are also the objects in which the reddening makes the smallest difference to the \Lbol. The uncertainty in \Lbol due to the uncertainty on $E(B\!-\!V)_{\rm int}$ varies from object to object. Typical values range from 0.03 dex (J1350$+$2652) to 0.16 dex (J2328$+$1500).

\begin{figure*}
	\centering
	\includegraphics[trim=1cm 2cm 1cm 2.5cm, clip=true, width = \textwidth]{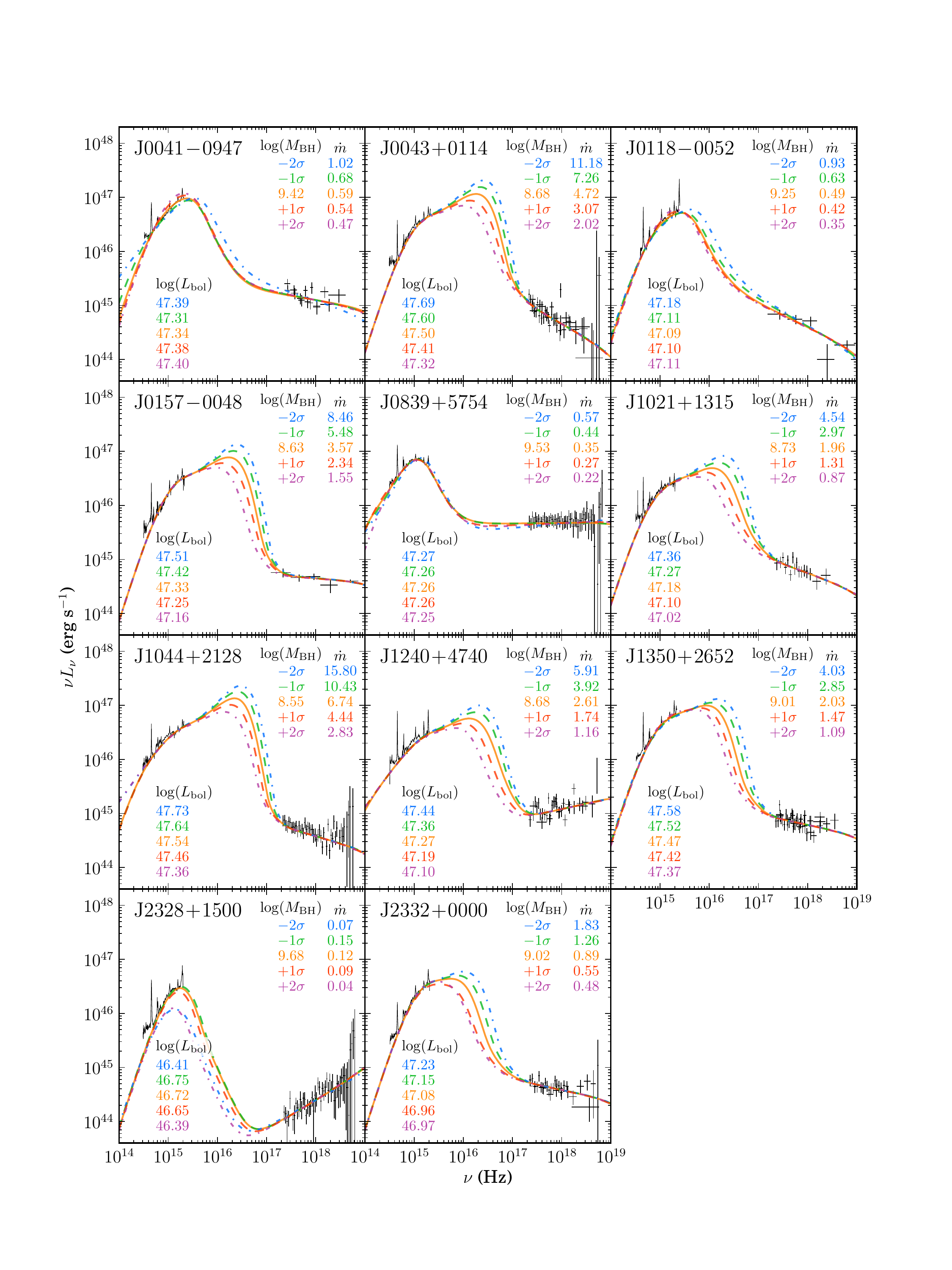}
	\caption{\small The SEDs for our sample, showing model variance with \M_BH. All models are at spin value, $a_*=0$. In black is the data from \xmmn (rebinned in three cases), SDSS and GNIRS/TSpec to which we fit our broad-band SED model. The model contains corrections for host galaxy extinction and soft X-ray absorption, as in Model 3 in Paper I, but here we use the best-fitting extinction curve (MW, LMC or SMC), rather than the MW curve in all cases. We have corrected the data for these sources of attenuation. The best fitting model, arising from our mean \M_BH estimate, is shown by the solid orange curve. Then we altered \M_BH by 1 and 2 $\sigma$ (see Table \ref{tab:sedmod} for \M_BH and errors) and remodelled the SED. The 1 $\sigma$ models are shown by the dashed curves and the 2 $\sigma$ models the dash-dot curves. The key model properties, with colours corresponding to the curves, are also given.}
	\label{fig:seds_mass}
\end{figure*}

The best fitting extinction curve in each object will be used henceforth. Our refined models are shown by the orange curves in Fig.\ \ref{fig:seds_mass}. We tabulate the key SED parameters and derived properties for these new models in Tables \ref{tab:sedpar} and \ref{tab:sedmod}.

\begin{table*}
	\caption{\small The optimum fitted parameters for the SED model fitted in Section \ref{subsec:intred}. The difference between this model and `Model 3' in Paper I is the extinction curve used. Whereas in Model 3 we used the MW curve in all objects, we noted that the reddening correction in J1044$+$2128 produced only a marginal fit to the data. Here we use the best fitting extinction curve out of the MW, LMC and SMC curves tested. In a few objects we have adjusted the optical/IR continuum points used in the fit from Paper I, hence the \rchi values are not directly comparable with Model 3 in that work. Uncertainties quoted are the 90 per cent confidence limits, as is conventional in X-ray astronomy, and are estimated using the Fisher matrix. As such, they are only indicative of the true measurement error. In objects where $f_{\rm SX}$ could not be constrained, we show the initial value (0.70) in brackets.}
	\small
	\centering
	\begin{tabular}{ccccccccccc}
		\hline
		Obj & ID & $N_{\rm H, \, int}$ & Extinct & $E(B-V)$   & $\dot{m}=$       & $r_{\rm cor}$ 
		& $r_{\rm out}$       & $\Gamma$         & $f_{\rm SX}$     & $\chi^2_{\rm red}$ \\
		&  & ($10^{22} \,$cm$^{-2}$)   & curve     & (mag)        & \Lbol$/$\LEdd
		& ($R_{\rm g}$)& ($R_{\rm g}$)&          &                  \\
		
		\hline
		1 & J0041$-$0947 & 0.0$\pm$0.2 & MW & 0.050$\pm$0.006 & 0.59$\pm$0.02 & 27$\pm$6 & 300$\pm$30 & 
		2.15$\pm$0.19 & 0.84$\pm$0.07 & 4.16 \\
		
		2 & J0043$+$0114 & 0.0$\pm$0.3 & LMC & 0.039$\pm$0.018 & 4.72$\pm$0.11 & 9.8$\pm$1.0 & 610$\pm$60 & 
		2.50$\pm$0.35 & (0.70) & 0.82 \\
		
		3 & J0118$-$0052 & 0.06$\pm$0.12 & MW & 0.025$\pm$0.014 & 0.49$\pm$0.02 & 25$\pm$11 & 240$\pm$40 & 
		2.44$\pm$0.14 & 0.77$\pm$0.14 & 1.92 \\
		
		4 & J0157$-$0048 & 0.19$\pm$0.10 & MW & 0.053$\pm$0.014 & 3.57$\pm$0.09 & 9.3$\pm$0.4 & 460$\pm$30 & 
		2.07$\pm$0.12 & (0.70) & 2.51 \\
		
		5 & J0839$+$5754 & 0.46$\pm$0.06 & MW & 0.066$\pm$0.003 & 0.349$\pm$0.005 & 80.9$\pm$1.6 & $>$1000 & 
		1.99$\pm$0.05 & 0.64 $\pm$ 0.03 & 1.62 \\
		
		6 & J1021$+$1315 & 0.2$\pm$0.2 & MW & 0.043$\pm$0.013 & 1.96$\pm$0.08 & 12$\pm$2 & 670$\pm$80 & 
		2.32$\pm$0.24 & (0.70) & 1.45 \\
		
		7 & J1044$+$2128 & 0.00$\pm$0.02 & SMC & 0.064$\pm$0.005 & 6.74$\pm$0.09 & 8.53$\pm$0.09 & $>$1000 & 
		2.24$\pm$0.05 & 0.7$\pm$0.4 & 1.63 \\
		
		8 & J1240$+$4740 & 0.00$\pm$0.11 & LMC & 0.039$\pm$0.011 & 2.61$\pm$0.10 & 14$\pm$7 & $>$1000 & 
		1.80$\pm$0.13 & (0.70) & 1.39 \\
		
		9 & J1350$+$2652 & 0.0$\pm$0.3 & MW & 0.034$\pm$0.007 & 2.03$\pm$0.05 & 9.7$\pm$0.4 & 440$\pm$20 & 
		2.19$\pm$0.14 & (0.70) & 1.98 \\
		
		10 & J2328$+$1500 & 0.00$\pm$0.15 & LMC & 0.16$\pm$0.04 & 0.122$\pm$0.007 & 14$\pm$3 & 40$\pm$3 & 
		1.48$\pm$0.09 & (0.70) & 1.63 \\
		
		11 & J2332$+$0000 & 0.0$\pm$0.3 & LMC & 0.08$\pm$0.03 & 0.890$\pm$0.10 & 10.9$\pm$1.5 & 176$\pm$12 & 
		2.18$\pm$0.14 & (0.70) & 0.61 \\
		
		\hline
	\end{tabular}
	\label{tab:sedpar}
\end{table*}

\begin{table*}
	\caption{\small The key properties of the various SED models, including BCs. We also show the \M_BH estimates calculated in Paper I. The errors on this value that we show here include the contribution from measurement error, and therefore are slightly larger than those shown in Table 3 of Paper I, where we tabulate the method error only. The error on \Lbol is estimated from the error on \mdot, and is indicative of the measurement error. The true error will be larger, due to additional contributions from \M_BH and $E$($B-V$)$_{\rm int}$.}
	\small
	\centering
	\begin{tabular}{ccccccccccc}
		\hline
		ID & $\log(M_{\rm BH} / M_{\odot})$           & $\log (L_{\rm bol})$       & $\log (L_{\rm 2-10 \, keV})$ & $\kappa_{\rm 2-10\,keV}$ 
		& $\log(\lambda L_{2500 \rm \mathring{A}})$& $\log(\nu L_{2\,\rm keV})$ & $\alpha_{\rm OX}$  
		& $\log(\lambda L_{5100 \rm \mathring{A}})$& $\kappa_{5100 \rm \mathring{A}}$ \\
		
		&                                          & [$\log$(erg s$^{-1}$)]     & [$\log$(erg s$^{-1}$)]       &  
		& [$\log$(erg s$^{-1}$)]                   & [$\log$(erg s$^{-1}$)]     & 
		& [$\log$(erg s$^{-1}$)]  &  \\
		\hline

		1 & 9.42$\pm$0.11 & 47.338$\pm$0.018 & 45.32 & 104  & 46.86 & 45.17 & 1.65 & 46.52 & 6.64  \\
		2 & 8.68$\pm$0.10 & 47.505$\pm$0.010 & 44.88 & 421  & 46.51 & 44.84 & 1.64 & 46.06 & 27.84 \\
		3 & 9.25$\pm$0.10 & 47.09$\pm$0.02   & 44.82 & 185  & 46.58 & 44.76 & 1.70 & 46.20 & 7.74  \\
		4 & 8.63$\pm$0.10 & 47.332$\pm$0.011 & 44.88 & 286  & 46.34 & 44.70 & 1.63 & 45.84 & 30.78 \\
		5 & 9.53$\pm$0.11 & 47.262$\pm$0.006 & 45.90 & 22.9 & 46.87 & 45.69 & 1.45 & 46.67 & 3.92  \\
		6 & 8.73$\pm$0.10 & 47.181$\pm$0.017 & 44.97 & 163  & 46.33 & 44.87 & 1.56 & 45.97 & 16.39 \\
		7 & 8.55$\pm$0.10 & 47.544$\pm$0.006 & 44.80 & 555  & 46.47 & 44.68 & 1.69 & 46.17 & 23.51 \\
		8 & 8.68$\pm$0.09 & 47.272$\pm$0.016 & 45.32 & 90.4 & 46.36 & 45.04 & 1.51 & 46.07 & 15.97 \\
		9 & 9.01$\pm$0.10 & 47.467$\pm$0.011 & 45.01 & 287  & 46.70 & 44.87 & 1.70 & 46.30 & 14.57 \\
		10 & 9.68$\pm$0.10 & 46.72$\pm$0.03  & 44.65 & 116  & 46.40 & 44.24 & 1.83 & 45.90 & 6.57  \\
		11 & 9.02$\pm$0.09 & 47.09$\pm$0.05  & 44.81 & 189  & 46.39 & 44.67 & 1.66 & 45.86 & 16.71 \\
		\hline
	\end{tabular}
	\label{tab:sedmod}
\end{table*}

\subsection{The Effect of Black Hole Mass on the SED} \label{subsec:mass}

In Paper I we commented upon the possible uncertainty in the SED model that may arise from the \M_BH estimate, in particular with regard to the SED peak position. To test this, we produce four new models for each AGN, with \M_BH varied by $\pm 1, 2 \sigma$ from its mean value. The modelling procedure is otherwise the same as described in Section \ref{subsec:intred}, with the same free and fixed parameters. The best-fit intrinsic extinction curve is used (Table \ref{tab:sedpar}). To avoid local minima in the fitting, and impartially test the total effect of altering \M_BH in each case, we apply the same modelling script in all cases, with the same initial values. Between models there can therefore be different values for all free parameters, including $E$($B-V$)$_{\rm int}$, \nh$_{\rm , int}$ and \rout . These may contribute to degeneracies between parameters, which it is also important to test for.

The total error on \M_BH is calculated by adding in quadrature the errors from the method dispersion and the measurement. These uncertainties are given in Table \ref{tab:sedmod}.

The resulting SEDs are presented in Fig.\ \ref{fig:seds_mass}, with accretion rates and \Lbol \ also given. For simplicity, only the dereddened/deabsorbed data for the mean \M_BH model is shown, hence models that do not appear to well describe the data are likely to have a different value of $E$($B-V$)$_{\rm int}$ or \nh$_{\rm ,int}$ (see J0041$-$0947 in Fig.\ \ref{fig:seds_mass} for a clear example of this variation).

It is clear that in objects with unconstrained SED peaks the difference is greatest. Reducing \M_BH produces an AD which peaks at higher energies, resulting in a larger \Lbol. In objects with well-constrained SED peaks, such as J0118$-$0052, this difference is smaller, and in J0839$+$5754 the difference is smallest of all, partly because the SED peak is dominated by the SX component, and therefore the peak temperature dependency on \M_BH is smaller. The intrinsic reddening value is consistent in all models, with all but J2328$+$1500 showing very little variation in optical/IR continuum slope. Degeneracy between the accretion rate and intrinsic reddening is evident in J2328$+$1500 however, with the \M_BH $\pm 2 \sigma$ models showing convergence to different optimum values of $E$($B-V$)$_{\rm int}$ (evinced by the lower SED for these models). It is encouraging that such an effect is only seen one object, and only when the \M_BH estimate is altered by $2 \sigma$ from the mean. In general the inherent uncertainty on \M_BH has only a small or predictable impact on the \Lbol , with a $\pm 0.1$ dex change in \M_BH propagating through to the \Lbol \ in all cases where the SED peak is unsampled.

\subsection{Exploring Spinning Black Holes} \label{subsec:spin}

So far, we have not investigated the effect of BH spin in the modelling, and find that all objects can be adequately fit with the $a_*$ spin parameter set to zero (\ie non-spinning), and the mass accretion rate left as a free parameter. However, this finding does not necessarily rule out high spin solutions, so here we will specifically explore $a_* \neq 0$ scenarios in our sample. With respect to the SED peak position, there is some degeneracy between the mass accretion rate and spin, as both affect the AD energy output and peak temperature. Therefore setting both parameters free in the fitting will not necessarily enable us to constrain the optimal spin value. Instead we repeat the SED fitting procedure for a range of additional $a_*$ values: 0.5, 0.8, 0.9 and 0.99. Other than these changes, the model fitting procedure is as described in Section \ref{subsec:intred}.

\begin{figure*}
	\centering
	\includegraphics[trim=1cm 2cm 1cm 2.5cm, clip=true, width = \textwidth]{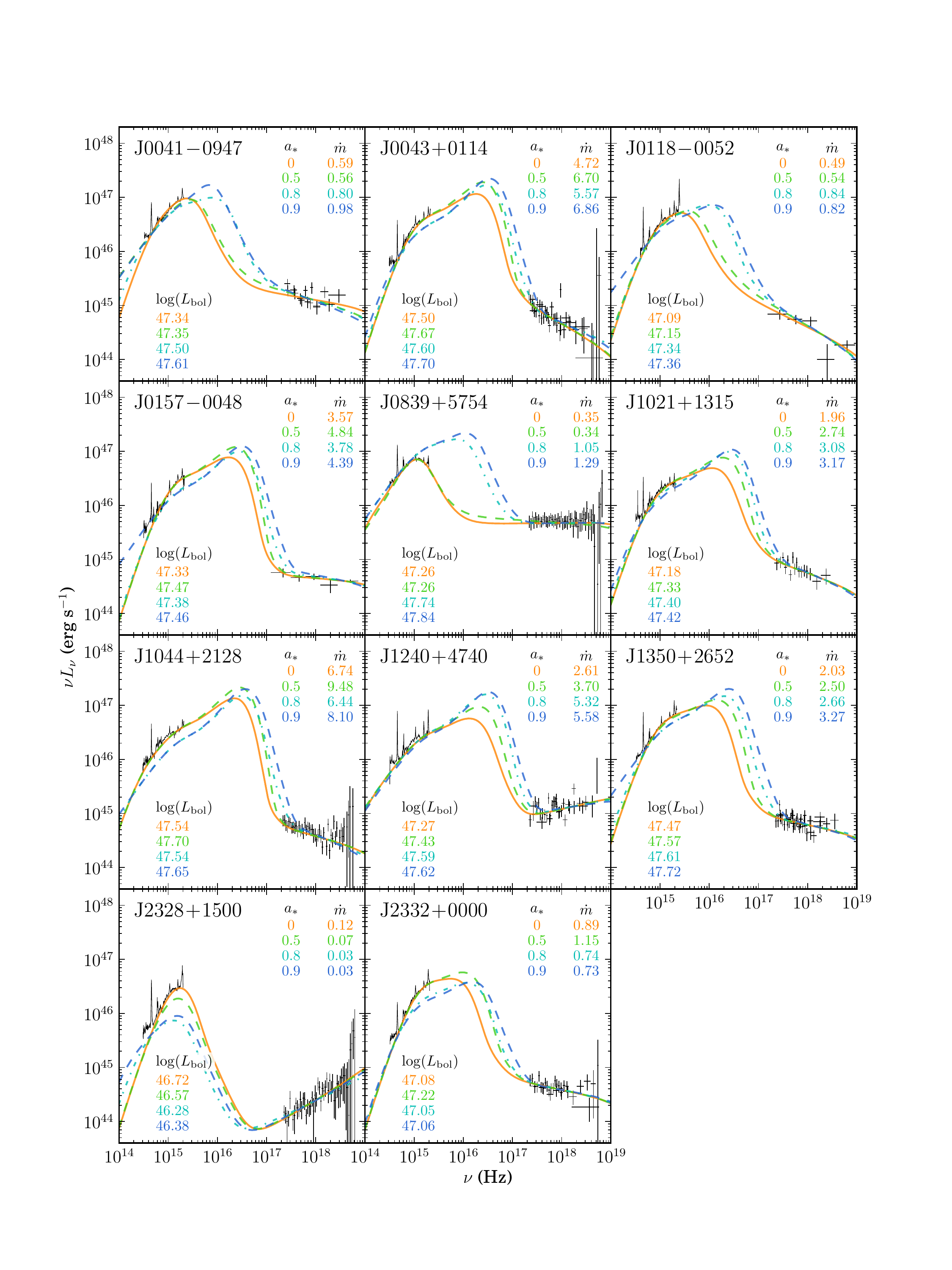}
	\caption{\small The SEDs for our sample, showing variation with the $a_*$ spin parameter. The data and orange curve are for the best fitting model with $a_* = 0$, and correspond to the identical model as shown by the orange curve in Fig. \ref{fig:seds_mass}. In each subsequent case, we fix $a_*$ at a higher value (0.5, 0.8, 0.9) and repeat the fitting procedure. We also tested a model with $a_* = 0.99$, but in all but one case, the resulting model either produced only a marginal fit, or a model in which nearly all of the AD energy is reprocessed in the Comptonised components, contrary to what is observed in local AGN. This result is dependent on relativistic corrections and inclination; see Section \ref{subsub:spin} for a discussion.}
	\label{fig:seds_spin}
\end{figure*}

In Fig.\ \ref{fig:seds_spin}, the SED models incorporating BHs with $a_*$ values of 0.5, 0.8, and 0.9 are shown alongside the $a_*=0$ model constructed in Section \ref{subsec:intred}. In $\sim 2/3$ of the sample we find that the moderate spin states ($a_*= 0.8, 0.9$) do not provide as good a fit to the data as the low spin states ($a_*=0, 0.5$), exhibited by the optical--NIR spectra (\eg J0839$+$5754) or by the X-ray spectra (\eg J1044$+$2128). Interpreting this result is complicated by the free parameters, in particular, the intrinsic attenuation properties of $E$($B-V$)$_{\rm int}$ and \nh$_{\rm , int}$, which are not immediately apparent in Fig.\ \ref{fig:seds_spin}. Three objects, J0041$-$0947, J1350$+$2652 and J2328$+$1500, show an improvement in the \rchi \ fitting statistic for the $a_*= 0.9$ model compared with $a_*= 0$, however the difference for the latter two is slight. This is discussed in Section \ref{subsub:spin}.

Using \optxagnf, we rule out the very highest spin states in our sample; for $a_* = 0.99$ our SED model breaks down in all but one (J2328$+$1500) case, producing SEDs that simply do not fit the data, or models where the PLT dominates the AGN energy output, in disagreement with previous work. The cause of this is that the energies resulting from these highest spin states cannot be redistributed in the Comptonised components. Given this, we do not plot the $a_* = 0.99$ model in Fig.\ \ref{fig:seds_spin}.

However, there are several important limitations of the AD model in \optxagnf that become significant here. \optxagnf assumes a fixed AD inclination to the observer of 60$^{\circ}$ and does not include relativistic effects. These corrections are fairly small when spin is low or zero, but become substantial as spin increases. This is discussed in Section \ref{subsub:spin}.

\subsection{Outer Accretion Disc Radius} \label{subsec:rout}

In all of the SED models we have produced so far, the outer AD radius (\rout) has been left as a free parameter. It has been suggested (\eg \citealt{goodman03}) that the AD extends out to a radius at which self-gravity causes it to break up, with the self-gravity radius, \rsg, depending on both \M_BH and accretion rate according to the following equation, given in \cite{laor89}:
\begin{equation} \label{eq:rout}
\left(\frac{r_{\rm sg}}{R_{\rm g}}\right) = 2150 \left(\frac{M_{\rm BH}}{10^9 M_{\odot}}\right)^{-2/9} \dot{m}^{4/9} \alpha^{2/9}
\end{equation}
where $\alpha$ is the ratio of viscous stress to pressure in the disc, fixed at a value of 0.1.

We explore this further by testing three different means of setting \rout:

\begin{itemize}
	\item[(1)] \rout free: this is the model described in Section \ref{subsec:intred}.
	\item[(2)] \rout fixed at \rsg.
	\item[(3)] \rout fixed at an arbitrary large value, as in \cite{jin12_1}.
\end{itemize}

\begin{figure}
\centering
\includegraphics[trim=0cm 0cm 0cm 0.7cm, clip=true, width = 0.5\textwidth]{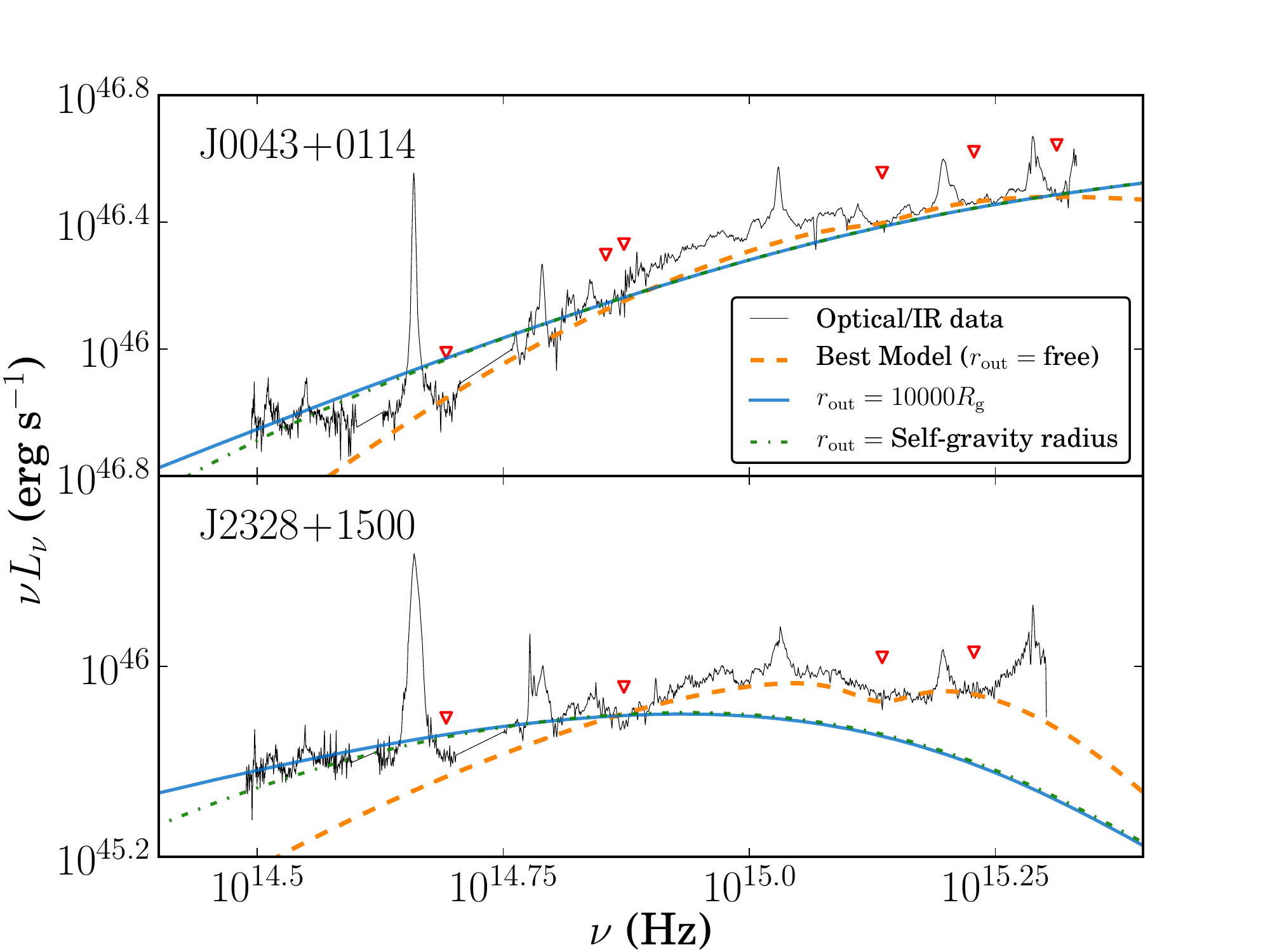}
\caption{\small Two example objects in which setting the \rout to an arbitrarily large value (10000 \Rg) or \rsg produces an inferior fit to the data. We show the intrinsically reddened spectra/SEDs here, whereas other figures show the `dereddened' (intrinsic) SED. The red markers show the continuum regions used for the fit.}
\label{fig:seds_rout}
\end{figure}

Based on both \rchi \ and visual inspection of the resulting models, in 8 of the 11 objects, when \rout is set to a large value or \rsg, the fit is poorer than those with \rout free. We show two examples in Fig.\ \ref{fig:seds_rout}. In the majority of cases, the model differences are confined mainly to the red part of the spectrum, (see J0043$+$0114 in Fig.\ \ref{fig:seds_rout}) as this is where the emission corresponding to the outer AD emerges. Notably, in J2328$+$1500 however, the difference between these models is also significant at short wavelengths, even though this emission originates from the inner AD regions. Here, the AD peak is predicted to fall short of the observed flux at short wavelengths, but the additional freedom in the model with \rout free allows it to converge to a solution with higher accretion rate and intrinsic extinction, resulting in a steeper intrinsic spectrum that requires a smaller \rout. In this test, BH spin was fixed at zero, and as noted in Section \ref{subsec:spin}, higher spin values also improve the fit to the short wavelength region of J2328$+$1500 without invoking intrinsic extinction (see also the discussion in Section \ref{subsub:spin}).

\begin{figure}
\centering
\includegraphics[trim=0cm 0.7cm 0cm 0.7cm, clip=true, width = 0.5\textwidth]{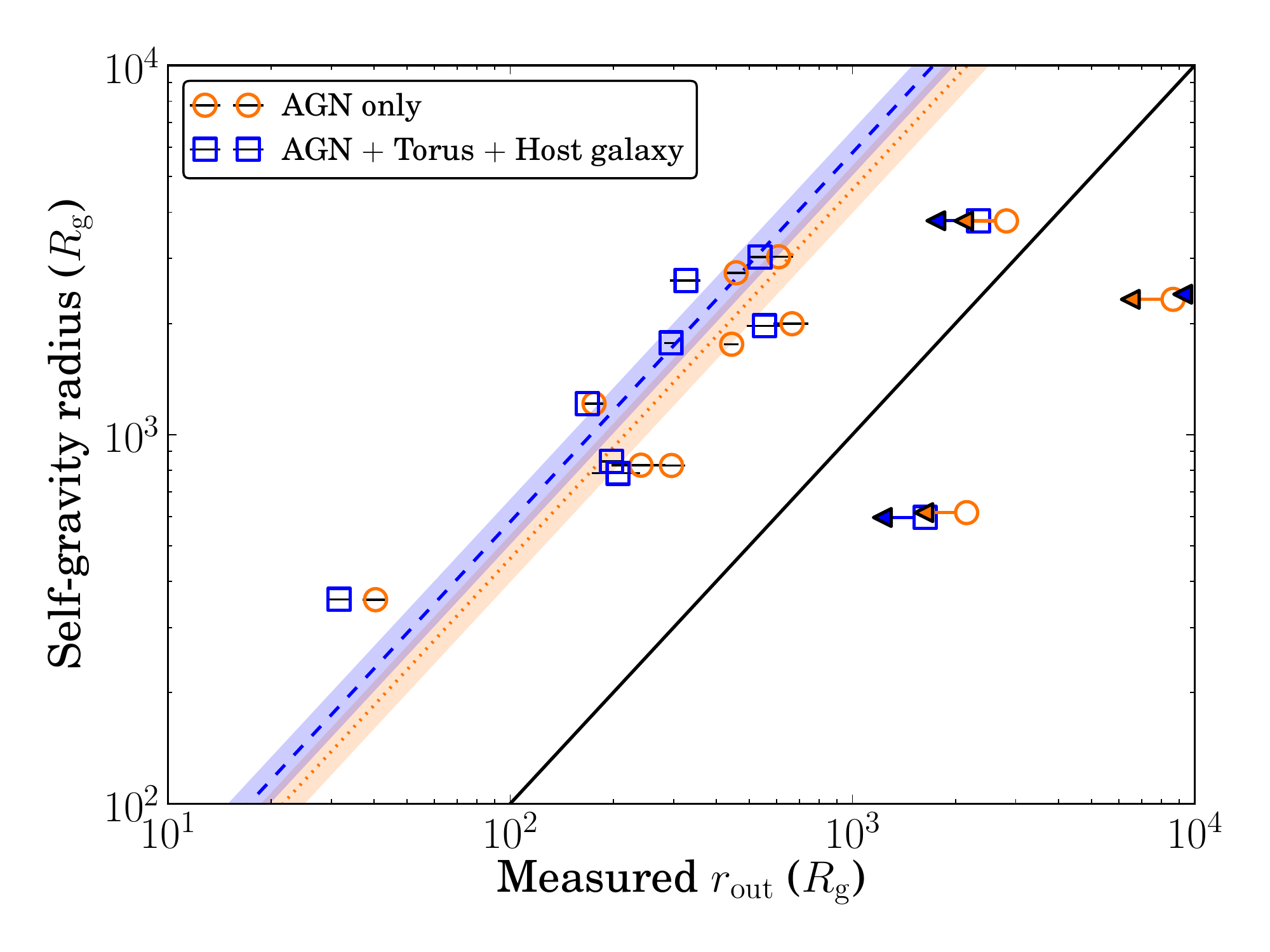}
\caption{\small Comparing \rout measurements from SED fitting with \rsg as given in equation \ref{eq:rout}. We show the three objects with unconstrained \rout as upper limits, and exclude these from the derived relations. The solid line represents unity, and the dotted and dashed lines represent the unity-gradient lines fitting the AGN only and AGN$+$Torus$+$Host galaxy model \rout values, respectively. The shaded regions show 1$\sigma$ error ranges.}
\label{fig:cf_rout}
\end{figure}

When we directly compare the \rout values we measure with those calculated from equation \ref{eq:rout} (Fig.\ \ref{fig:cf_rout}), we do see an increasing trend with the \rsg, but offset from unity, suggesting that self-gravity may play a role in setting \rout, but is not the only contributing factor. Also shown are the \rout values derived in the following section, where we include model components for the torus and host galaxy (Section \ref{sec:torus}). Our derived relations are:
\begin{equation}
\log(r_{\rm out}) = \log(r_{\rm sg}) - (0.66\pm0.06)
\end{equation}
for the AGN only model, and:
\begin{equation}
\log(r_{\rm out}) = \log(r_{\rm sg}) - (0.76\pm0.06)
\end{equation}
for the model including torus and host galaxy components.

\section{Torus and Host Galaxy} \label{sec:torus}

\begin{figure*}
	\centering
	\includegraphics[trim=1cm 1.5cm 1cm 2.5cm, clip=true, width = \textwidth]{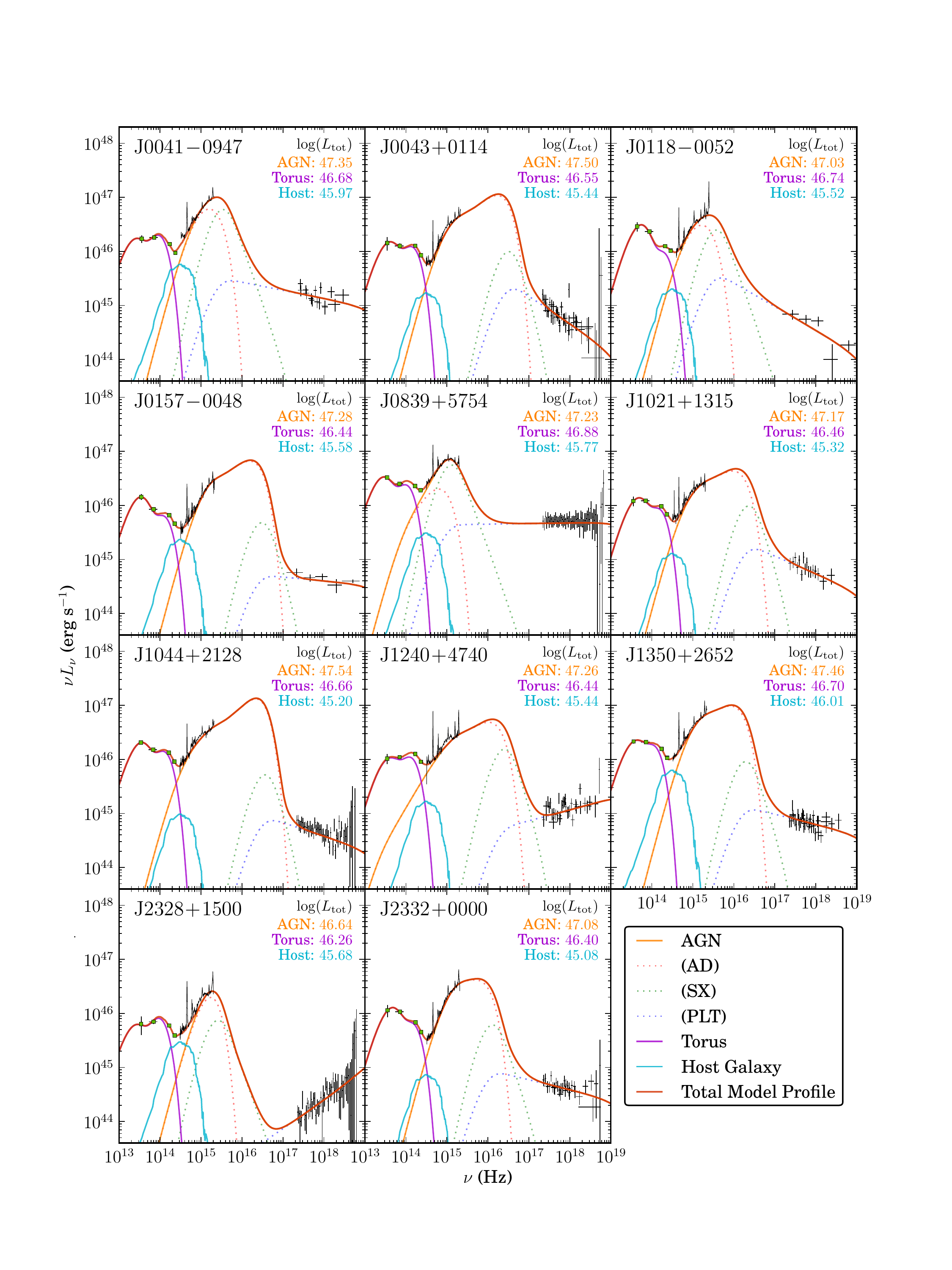}
	\caption{\small The full mid-IR to X-ray SEDs for our sample. The green squares show the \textit{WISE} photometry data, and the optical--NIR and X-ray data are once again shown in black. The orange, light blue and purple curves show the contributions from the AGN, host galaxy and dusty torus components respectively. The torus comprises two blackbody components, and the host galaxy is a 5 Gyr elliptical template.}
	\label{fig:seds_torus}
\end{figure*}

\begin{table*}
	\caption{\small Luminosities and temperatures of the different components in our sample.}
	\small
	\centering
	\begin{tabular}{ccccccccc}
		\hline
		
		Name & $\log(L_{\rm bol})$         & $\log(L_{\rm Torus})$       & $\log(L_{\rm Host})$        & 
		$\frac{L_{\rm 1\mu m, host}}{L_{\rm 1\mu m, total}}$ & $T_{\rm hot}$ & $T_{\rm warm}$ & $C_{\rm hot}$ & $C_{\rm warm}$  \\
		& [$\log(\rm erg \: s^{-1})$] & [$\log(\rm erg \: s^{-1})$] & [$\log(\rm erg \: s^{-1})$] & 
		& [K]           & [K]            &  [\%]         & [\%] \\
		\hline
		J0041$-$0947 & 47.35$\pm$0.02 & 46.7$\pm$0.3   & 45.97$\pm$0.09 & 0.46 & 1090$\pm$160  & 300$\pm$1400 & 
		12.7 & 8.7 \\
		J0043$+$0114 & 47.50$\pm$0.02 & 46.55$\pm$0.02 & 45.4$\pm$0.1   & 0.28 & 1630$\pm$60   & 460$\pm$60   &
		5.5 & 5.7 \\
		J0118$-$0052 & 47.03$\pm$0.03 & 46.74$\pm$0.02 & 45.5$\pm$0.4   & 0.19 & 2170$\pm$170  & 650$\pm$ 60  &
		12.8 & 38.2 \\
		J0157$-$0048 & 47.28$\pm$0.02 & 46.44$\pm$0.02 & 45.58$\pm$0.08 & 0.62 & 1410$\pm$90   & 400$\pm$30   &
		4.8 & 9.8 \\
		J0839$+$5754 & 47.23$\pm$0.07 & 46.9$\pm$0.6   & 45.8$\pm$0.3   & 0.17 & 1200$\pm$1300 & 300$\pm$500  &
		20.5 & 24.3 \\
		J1021$+$1315 & 47.17$\pm$0.02 & 46.46$\pm$0.02 & 45.3$\pm$0.2   & 0.24 & 1710$\pm$70   & 540$\pm$ 50  &
		8.4 & 11.4 \\
		J1044$+$2128 & 47.54$\pm$0.01 & 46.66$\pm$0.02 & 45$\pm$2       & 0.13 & 1450$\pm$50   & 410$\pm$30   & 
		5.7 & 7.3 \\
		J1240$+$4740 & 47.26$\pm$0.04 & 46.4$\pm$0.5   & 45.4$\pm$0.1   & 0.21 & 1500$\pm$70   & 400$\pm$1700 &
		8.6 & 6.5 \\
		J1350$+$2652 & 47.46$\pm$0.01 & 46.70$\pm$0.03 & 46.01$\pm$0.03 & 0.61 & 1270$\pm$70   & 440$\pm$50   &
		8.8 & 8.3 \\
		J2328$+$1500 & 46.64$\pm$0.02 & 46.3$\pm$0.3   & 45.68$\pm$0.03 & 0.72 & 1130$\pm$90   & 300$\pm$800  &
		26.2 & 15.7 \\
		J2332$+$0000 & 47.08$\pm$0.02 & 46.40$\pm$0.02 & 45.1$\pm$0.2   & 0.22 & 1650$\pm$80   & 540$\pm$60   &
		7.7 & 13.1 \\
		\hline
		
	\end{tabular}
	\label{tab:torho_props}
\end{table*}

In Paper I we discussed the potential contribution of the host galaxy to the total SED. This may be manifest in the `red excess' we observe in nearly all objects, redward of the \Ha emission line. The \cite{jin12_1} sample was at $z<0.3$, and was therefore of lower average luminosity than our sample. As such, many of their AGN exhibited significant host galaxy contamination in the optical spectral continuum. In general, for AGN at $z\gtrsim0.5$ the host galaxy flux is assumed to be insignificant (\eg \citealt{shen11}), and indeed we concluded in Paper I that for our least luminous source (J2328$+$1500), the maximum possible contribution to the SED peak (at $\sim 2000 \; \rm \mathring{A}$) was $\sim 1$ per cent, which increased to $\sim 50$ per cent at a wavelength of $\sim 1 \mu$m.

Since this object also hosts the most massive BH of our sample, it is expected to exhibit the largest contamination by the host galaxy, based on the \M_BH--bulge mass relationship (see Section \ref{sec:intro}). We therefore concluded that the host galaxy contribution to the total SED continuum is small in all objects. Nonetheless, the red excess and \textit{WISE} photometry show evidence for a dusty torus component, possibly including flux from the host galaxy. Thus, as the final refinement of our SED modelling we now include SED components for both the torus and host galaxy, in order to fit the spectral data redward of \Ha (regions 10--15 in Table \ref{tab:cont_wav}), and the \textit{WISE} photometry.

In practice, the torus is known to have a complex SED, comprising blackbody emission from the (possibly clumpy) hot dust, and emission/absorption from atomic/molecular transitions, including polyaromatic hydrocarbons related to star formation \citep{schweitzer06}. However, due to data limitations, we will model the torus with only two blackbody components, hereafter referred to as `hot' and `warm'. The temperature of the hot component, $T_{\rm hot}$, informs us of the composition of the dust grains that form the torus. Silicate grains sublimate at temperatures above \mbox{$\sim 1500$ K,} whereas graphitic grains can survive up to $\sim 1800 - 2000$ K (\eg \citealt{barvainis87}, \citealt{mor09}, \citealt{netzer15}). In this respect our approach is similar to that employed by \cite{mor11}, who modelled a single, hot graphitic dust component in a large sample of AGN, and \cite{landt11}, who used blackbody models of the hot dust in their sample of AGN. \cite{kirkpatrick15} also modelled combined blackbody components to represent the warm and cold dust in their sample of luminous IR star-forming galaxies and AGN.

In Paper I we tested two models of the host galaxy, that of a 5 Gyr elliptical galaxy and a starburst galaxy (represented by M82) with a stronger SED contribution at UV wavelengths. We extracted these galaxy templates from \cite{polletta07}. Practically, the difference between the two templates was small, as UV flux is dominated by AD emission. Based on the \M_BH--bulge relationship, we expect our sample of AGN to be hosted by massive elliptical galaxies, and we will therefore use the 5 Gyr template of \cite{polletta07} only in this work.

We fit the SEDs in \xspec again, fixing the X-ray part of the spectrum to values calculated in Section \ref{subsec:intred}. The full mid-IR to X-ray SEDs, including the torus and host galaxy components, are shown in Fig.\ \ref{fig:seds_torus}. We tabulate the key parameters in Table \ref{tab:torho_props}. Dust covering factors are calculated for both the hot and warm torus components using the formula $C=L_{\rm dust} / L_{\rm bol}$, where $L_{\rm dust}$ and $L_{\rm bol}$ are the luminosities of the hot/warm dust, and AGN, respectively.

\section{Optical Variability} \label{sec:variability}

In Paper I we briefly discussed the possible causes of variability in our AGN sample. Since our multiwavelength data are collected non-contemporaneously, it is important to look for evidence of variability between optical and NIR observations. We only identified such evidence in J0041$-$0947. In this case, for the SED fitting we normalised the NIR spectrum to the level of the optical spectrum. We also presented all available optical spectra from SDSS/BOSS, as five objects in our sample had optical spectra taken on multiple epochs. Comparing these spectra, there appears to be some small ($\sim 20$ per cent) variability in three of the objects observed more than once.

\begin{figure}
	\centering
	\includegraphics[width = 0.5\textwidth]{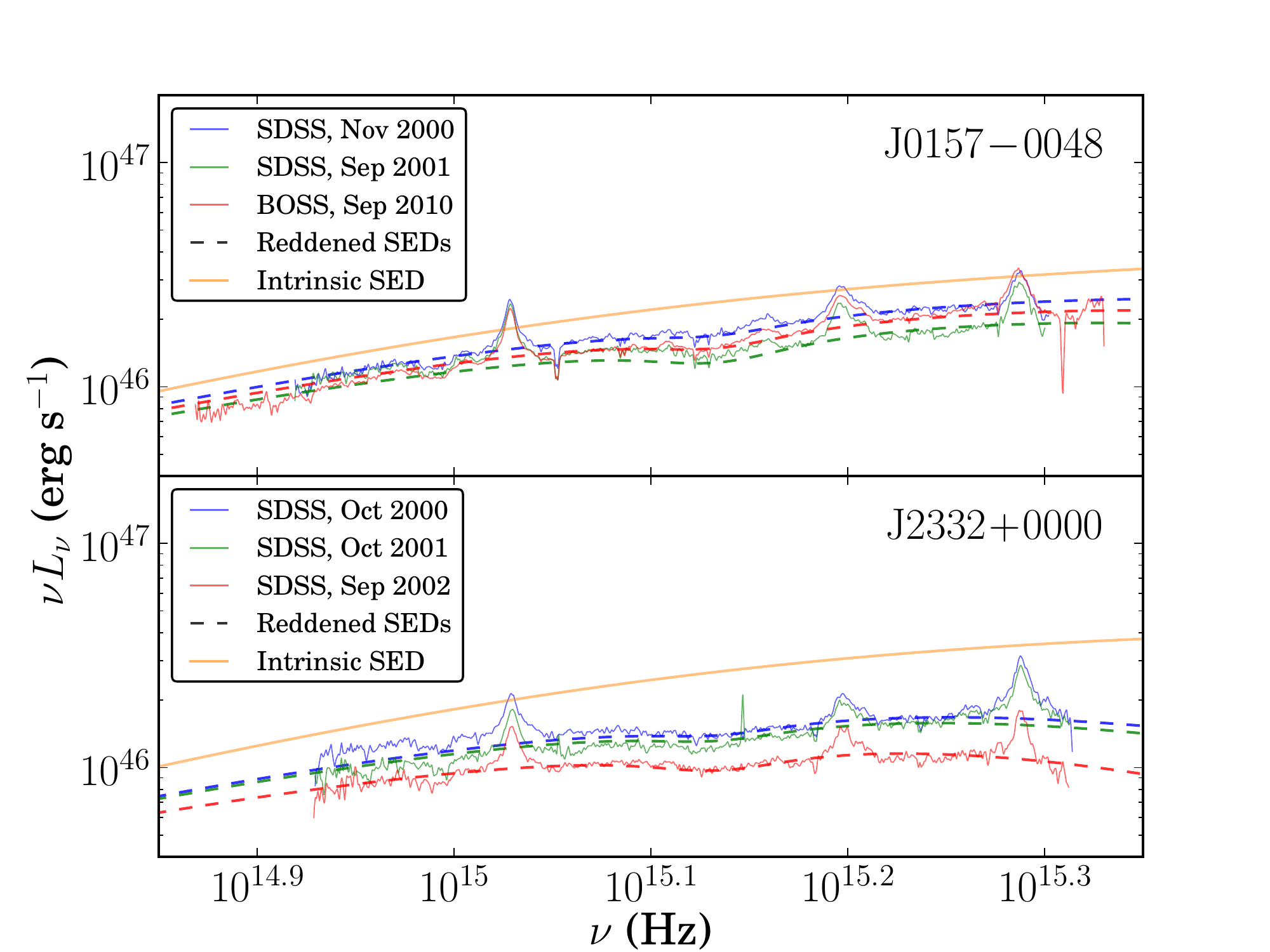}
	\caption{\small Examples of the `variable' objects in our sample. Five objects were observed on multiple occasions in the optical by SDSS/BOSS, and of these, three showed some evidence for variability. Modelling this as an effect of changing extinction produces a first order correction for this variability. The intrinsic SED therefore does not change between observations, but the observed SED does change. In these two examples, we show the data from each different epoch in different colours (with the same colour scheme as in Paper I), and the attenuated model as a correspondingly coloured dashed line. The difference in $E$($B-V$)$_{\rm int}$ between Oct 2000 and Sep 2002 in J2332$+$0000 is 0.043 mag. All spectra have been corrected for Milky Way extinction.}
	\label{fig:var}
\end{figure}

We examined the possible cause of these variations, using our SED model. There are few specific AGN properties that can change on timescales of a few years; \M_BH and spin are fixed and changes in the mass accretion rate are physically limited by the viscous timescale -- the characteristic time taken for mass to flow through the disc. This timescale for the BH masses in our sample is likely to be of the order of hundreds of years even in the innermost regions of the AD \citep{czerny04}, although AGN variability is frequently noted that occurs faster than the viscous timescale (\eg \citealt{denney14}, \citealt{lamassa15}). This could be due to reprocessing of SX emission in the AD, as PLT emission is generally too weak to have a significant effect in most of our objects \citep{gardner15}. The intrinsic reddening we model could change if the extinguishing dust is `clumpy' in nature, as is thought to be the case for the torus (\citealt{risaliti05}), thus presenting a constantly changing $E$($B-V$)$_{\rm int}$ parameter.

In order to explore the effects of changing the obscuration, we take our best-fitting SEDs, and attempt to fit any other epochs of data available by adjusting only the $E$($B-V$)$_{\rm int}$ parameter. In Fig.\ \ref{fig:var} two examples of the variable AGN are shown, together with models that result from changing $E$($B-V$)$_{\rm int}$. To first order we find that the variability we observe between observations could be attributed to changing extinction but the optical spectrum alone covers too short a wavelength range to test this hypothesis effectively -- there are often only 2--3 emission free regions in the optical spectra to which we fit the SED model. A stronger test of the changing properties that are responsible for such variability would require simultaneous optical--NIR data from multiple epochs, as this would provide the data coverage required to model the AD robustly. It may be that changes in the accretion rate must also be considered to fully parameterise the observed spectral variability. All such tests also require accurate flux calibration; uncertainties in the SDSS flux calibration may also contribute to observed apparent variability.

\section{Optical--NIR Spectral Decomposition} \label{sec:spec_decomp}

Our physical model of the underlying AGN continuum now enables us to perform a complete decomposition of the optical--NIR spectrum, including the contribution from the BLR. The BLR is thought to lie between the AD and torus (\eg \citealt{antonucci93}, \citealt{beckmann12}, \citealt{czerny15}), with electron transitions in partially-ionised gas giving rise to many emission features that are Doppler-broadened by the rapid orbit of this gas around the BH. Our sample is too small to improve on the emission line correlations extensively studied by \eg \cite{shen12}, \cite{denney13}, \cite{karouzos15}, however, since such studies generally use power-law continuum models, it is desirable to have a better understanding of the true continuum, especially as this continuum forms the basis of many virial \M_BH estimators. In particular, the Balmer continuum lies underneath the \Mgii feature, but due to additional contamination by \Feii, this can be difficult to deconvolve, particularly when considering only a limited wavelength range on either side of the \Mgii lines.

Our spectral model will include models of the isolated emission lines, and a `pseudo-continuum' which includes blended line emission as well as true continuum contributions.

The emission lines are modelled as superpositions of Gaussians. Whilst this is an approximation to the true emission line shape, it provides a versatile and widely adopted means of characterising the emission lines (see \eg \citealt{greene07}, \citealt{dong08}, \citealt{wang09}, \citealt{matsuoka13}, and also \citealt{assef11} and \citealt{park12} and references therein for examples of alternative models using Gauss-Hermite polynomials).

We use the following components for the emission lines:

\begin{itemize}
	\item[i.]{\Ha $\lambda 6563$ is fitted with two broad components (or one broad, and one `intermediate'). As in Paper I, for objects that show strong narrow \Oiii, we include a third, narrow Gaussian component, locked in velocity width and wavelength to the strong, narrow \Oiii member.}
	\item[ii.]{\Hb $\lambda 4861$ is fitted with an equivalent profile to \Ha, with only the normalisation as a free parameter.}
	\item[iii.]{\Oiii $\lambda 4959\rm ,\, 5007$ is a doublet. We fit each member with two Gaussians, or one Gaussian for objects showing particularly weak \Oiii emission (J0043$+$0114, J0157$-$0048, J1021$+$1315, J1044$+$2128 and J1240$+$4740).}
	\item[iv.]{\Hg $\lambda 4340$ is fitted in the same manner as \Hb.}
	\item[v.]{\Mgii $\lambda 2798$, \Ciii $\lambda 1908$, \Civ $\lambda 1549$ and \Lya $\lambda 1216$ are modelled with two broad Gaussian components each. We do not include narrow components for these lines as in general there is no statistical justification for a third component. \Lya is only covered in J0118$-$0052. We do not attach a physical significance to the two components in any of these lines. For instance, \Mgii is a doublet, but we do not model it as such as the line splitting is too small to be significant.}
	\item[vi.]{\Hei $\lambda 5876$, \Hd $\lambda 4102$, \Neiii $\lambda 3869$, \Oii $\lambda 3729$, \Neiv $\lambda 2422$, \Cii $\lambda 2326$, \Aliii $\lambda 1855$, \Heii $\lambda 1640$, \Siiv $\lambda 1394$ (may include \Oiv) and \Oi $\lambda 1305$ (may include \Siii) are all modelled for completeness with a single Gaussian component, though most are very faint in our spectra, so we freeze their wavelengths to literature values \citep{vandenberk01}.}
\end{itemize}

The pseudo-continuum comprises the following components:

\begin{figure*}
	\centering
	\includegraphics[trim=1cm 2cm 1cm 2.5cm, clip=true, width = \textwidth]{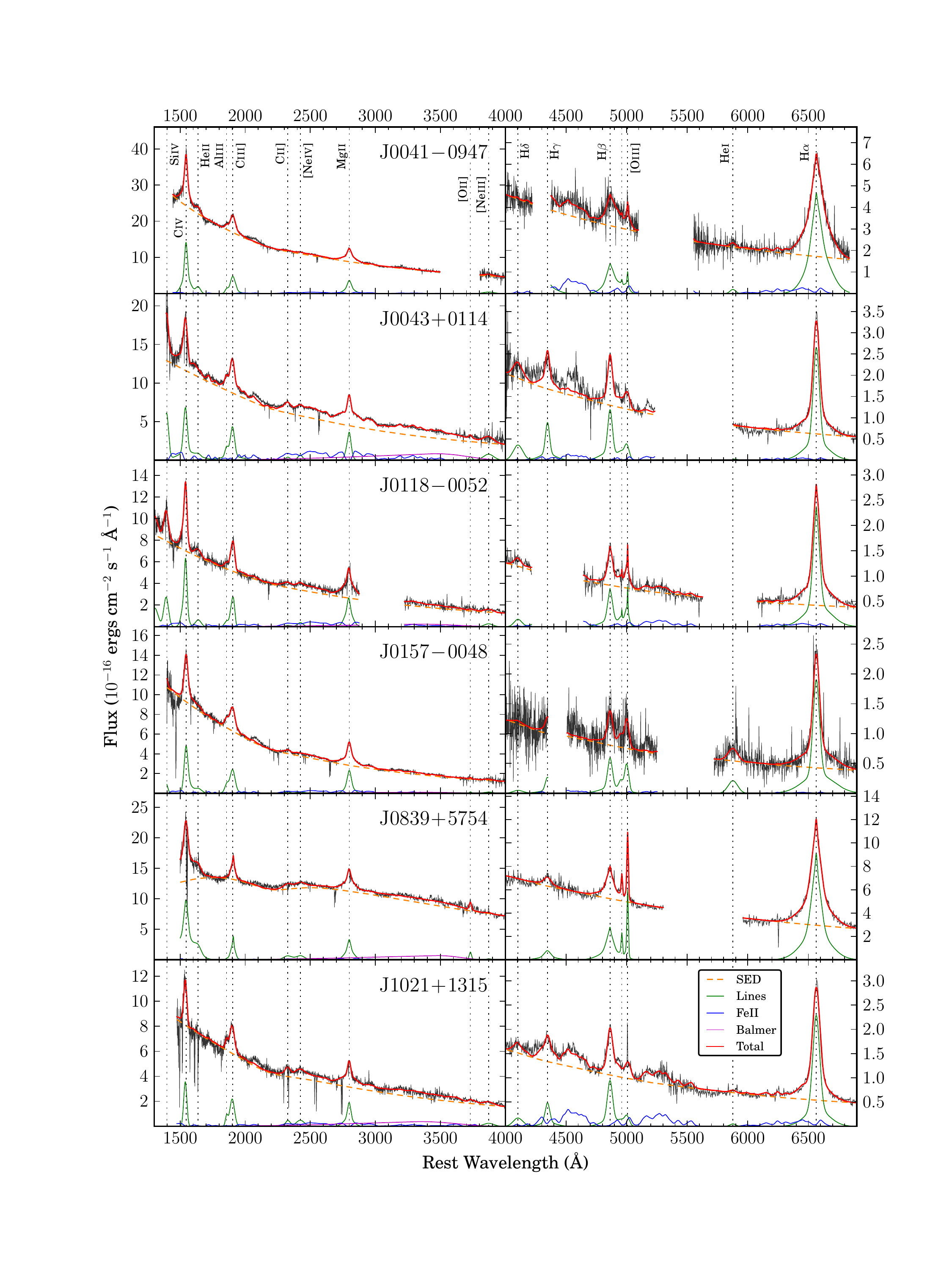}
	\caption{\small Optical--NIR spectral decomposition using our SED model together with models of the Balmer continuum, blended \Feii emission, and emission lines. Line identifications are shown in the top panels. The right panels are magnified to more clearly show the Balmer region.}
	\label{fig:specdecomp1}
\end{figure*}

\begin{figure*}
	\centering
	\includegraphics[trim=1cm 5.7cm 1cm 2.5cm, clip=true, width = \textwidth]{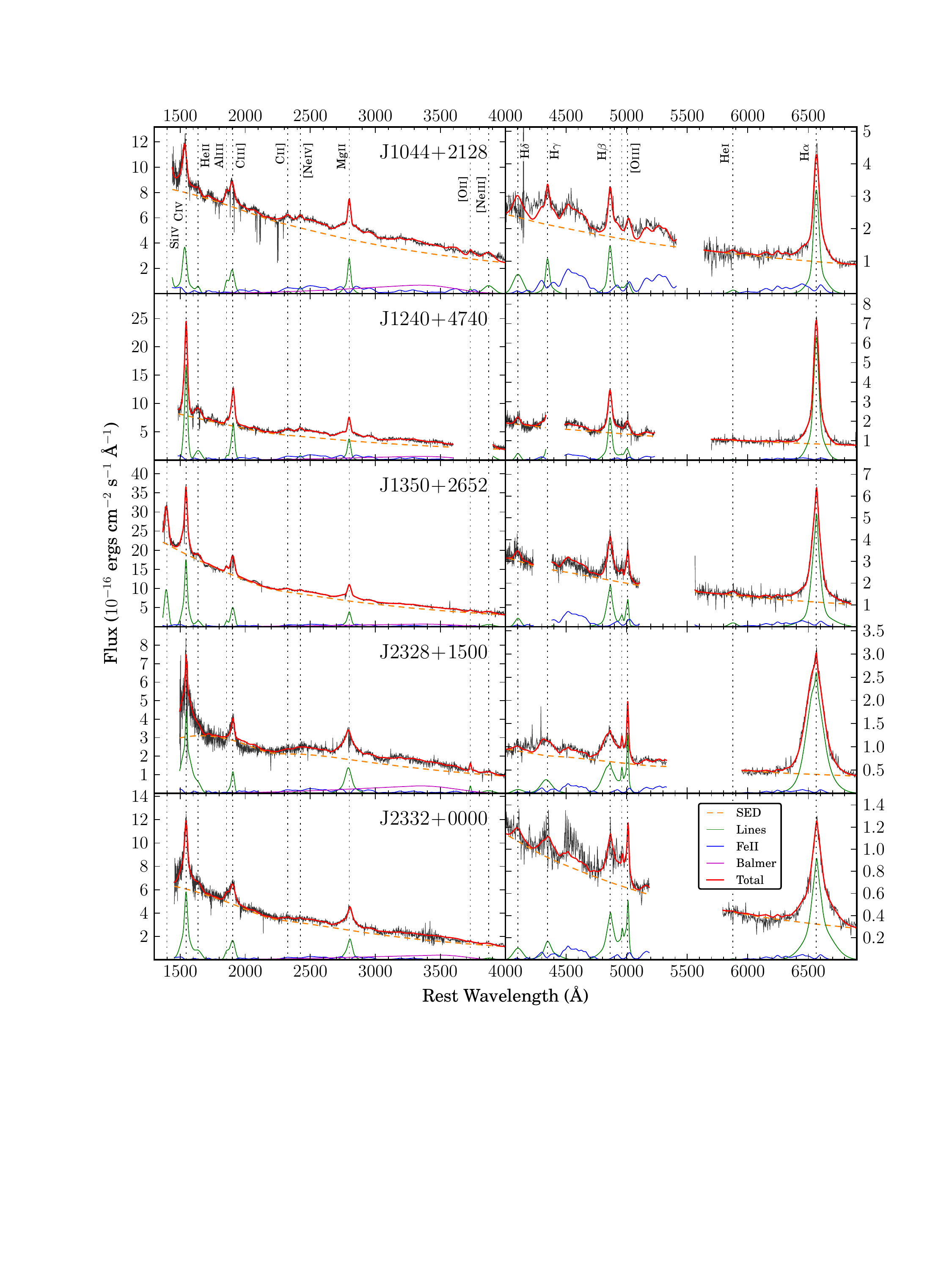} 
	\caption{\small Continuation of Fig.\ \ref{fig:specdecomp1}}
	\label{fig:specdecomp2}
\end{figure*}

\begin{itemize}
\item[i.]{\textbf{\textsc{optxagnf} continuum}: We use the model constructed in Section \ref{sec:torus}, as it incorporates the host galaxy and dust components. An exception is J0041$-$0947, for which we adopt the model with a BH spin parameter of $a_*=0.9$. As noted in Section \ref{sec:sedmodel}, this is the one object in our sample where we see a significant improvement in the continuum fit when we introduce a spinning BH ($\chi_{\rm red} = 4.16, 0.92$ for $a_* = 0, 0.9$). We allow some freedom in the normalisation; if the continuum regions chosen for SED model-fitting are marginally contaminated by an emission component, the true continuum could be below that we calculate.}
\item[ii.]{\textbf{Balmer continuum}: We employ the following model of the Balmer continuum (\eg \citealt{grandi82}, \citealt{jin12_1}):
\begin{equation}
F_{\nu,\rm BC} = F_{\nu,\rm BE} \, e^{-h(\nu - \nu_{\rm BE})/(kT_{\rm e})} \:\:\:\:\:\:\: (\nu \geqslant \nu_{\rm BE})
\end{equation}
where $\nu_{\rm BE}$ and $F_{\nu,\rm BE}$ are the frequency and flux density at the Balmer edge, respectively. We convolve this model with a Gaussian to account for Doppler-broadening associated with intrinsic velocity dispersion in the hydrogen emitting gas. $\nu_{\rm BE}$ (initial value of \mbox{3646 \AA}, \citealt{jin12_1}), the temperature, $T$, and the width of the convolving Gaussian are free parameters.}
\item[iii.]{\textbf{Blended Fe\textsc{ii} emission}: To model the ubiquitous, blended permitted \Feii emission seen throughout the optical--NIR AGN spectrum, we use two empirical templates, derived from the Type 1 AGN I Zwicky 1. These templates come from \cite{veroncetty04} for the (rest frame) optical range and \cite{vestergaard01} in the UV. We use the theoretical \Feii emission template of \cite{verner09} for the 3100 -- 3500 \AA \ gap between these. The templates are convolved with a Gaussian to incorporate velocity broadening, and normalised independently in the optical and UV. The normalisation and Gaussian width are free parameters.}
\end{itemize}

After fitting the pseudo-continuum, we then fit the emission lines systematically. All fitting is performed by custom \python scripts, using the Levenberg-Marquardt algorithm provided in the {\sc lmfit} package\footnote{http://lmfit.github.io/lmfit-py/}. To estimate measurement errors, we use a Monte Carlo method, where 100 different realisations of the spectra are created using the mean (measured) flux density and standard error at each pixel, and refitting the model from scratch. The central 68 per cent of the resulting value distribution for any given property then provides an estimation of the measurement error. In these `mock' spectra, the artificial `noise' is added to already noisy spectra, but to an approximation, this method will give a good representation of the true errors.

We show the model fits resulting from our spectral decomposition in Figs.\ \ref{fig:specdecomp1} and \ref{fig:specdecomp2}.

\section{Discussion} \label{sec:discussion}

\subsection{Refined SED Model Properties} \label{subsec:sed_prop}

\begin{figure}
	\centering
	\includegraphics[trim=0cm 0cm 0cm 0.78cm, clip=true, width = 0.5\textwidth]{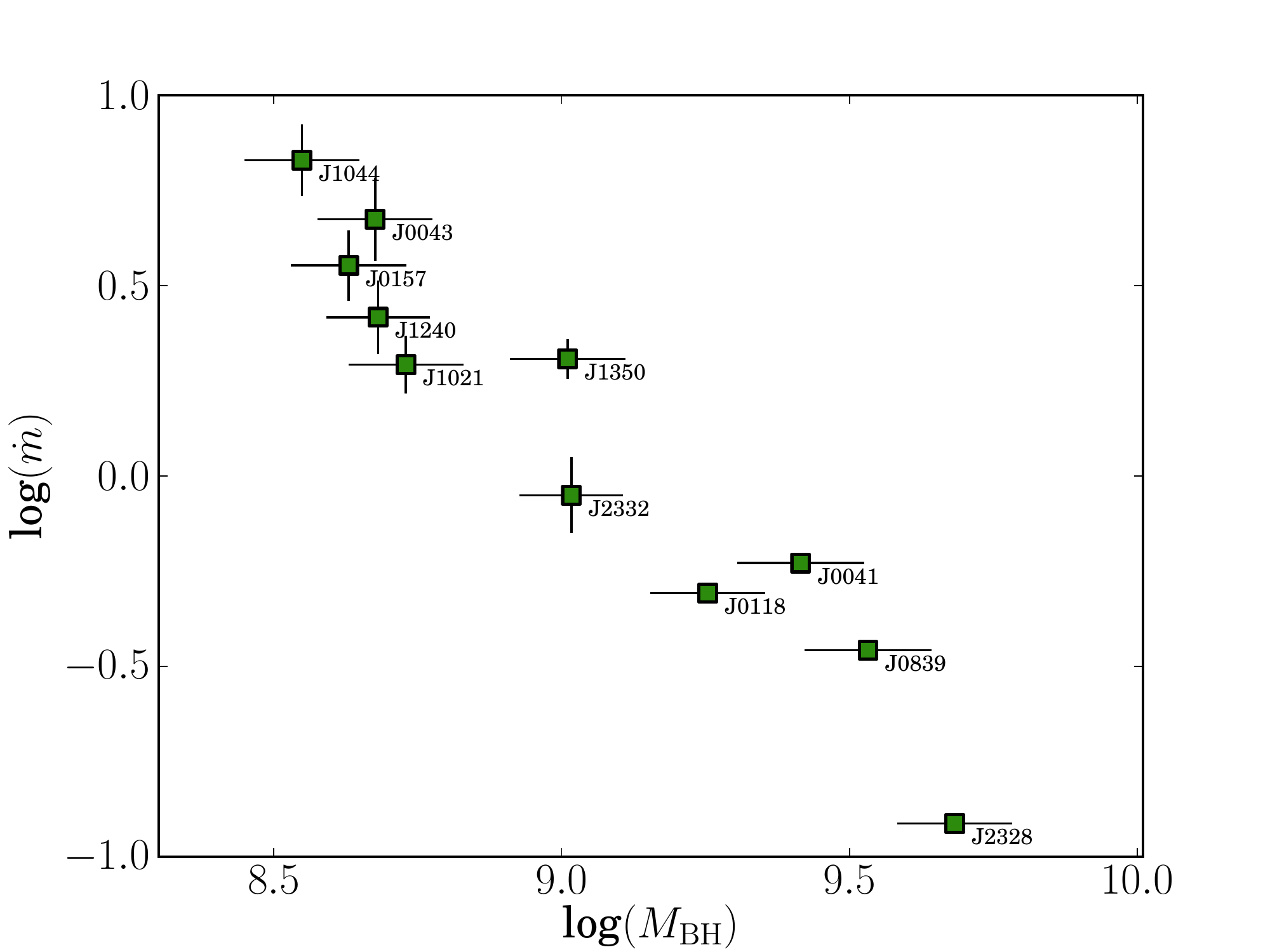}
	\caption{\small Dependence of \mdot on \M_BH. This highlights selection effects for our sample.}
	\label{fig:mbh_mdot}
\end{figure}

Our data are well described by the physically-motivated SED continuum and emission line models we have built. We refine the SED models of Paper I in Section \ref{subsec:intred} by improving our treatment of intrinsic reddening. We will first discuss the properties of these models, and compare them to similar studies. We note there is an anti-correlation between \M_BH and \mdot (Fig.\ \ref{fig:mbh_mdot}), likely because our sample is selected from a small redshift range, are of similar flux, and are therefore of comparable luminosity. Previous work, such as \cite{vasudevan07}, \cite{davis11}, \cite{jin12_3} have suggested that it is \mdot that more strongly governs the observed SED properties, including the X-ray photon index and BCs.

\begin{figure}
	\centering
	\includegraphics[trim=0cm 0.78cm 0cm 0cm, clip=true, width = 0.5\textwidth]{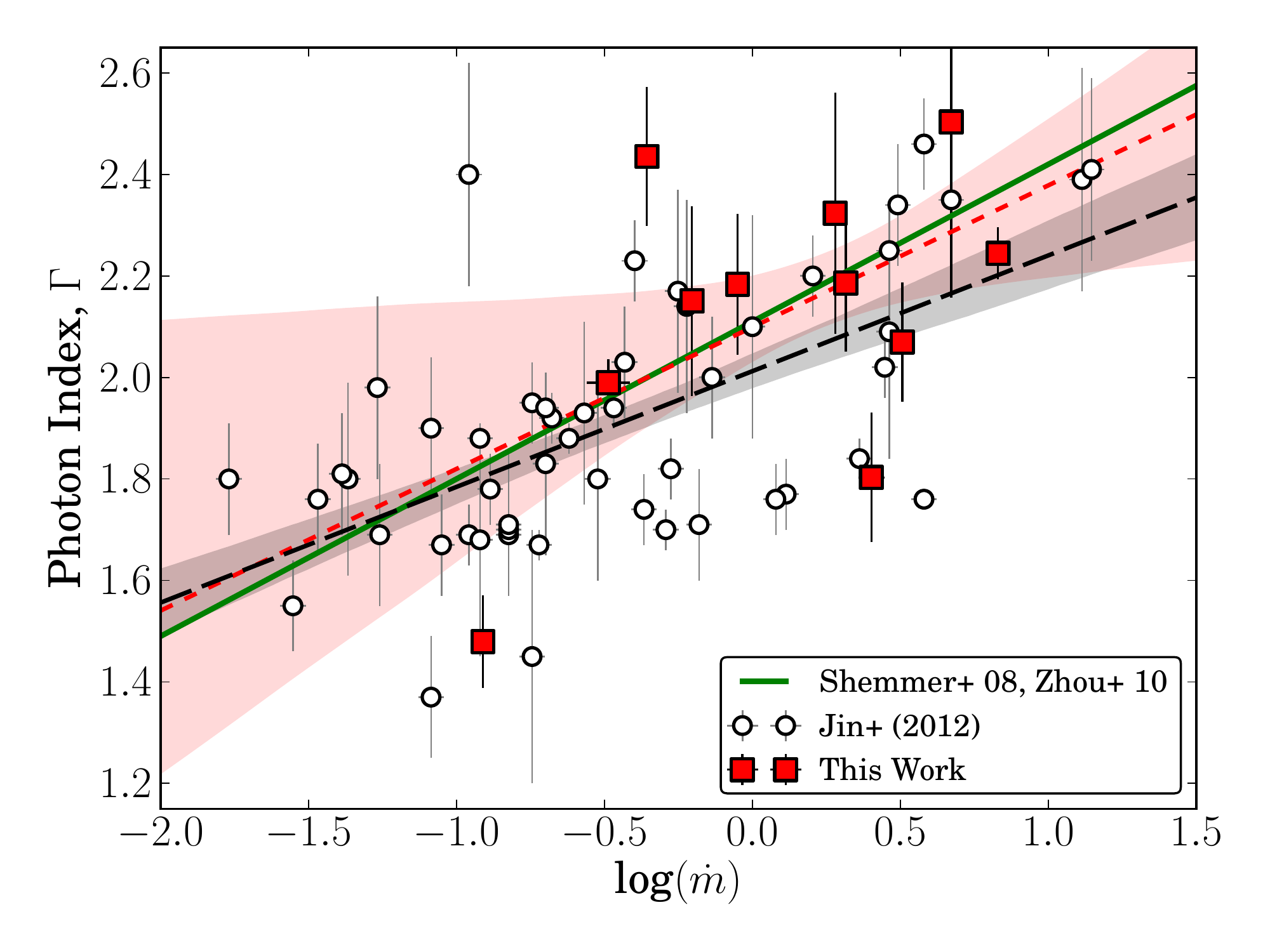}
	\caption{\small The photon index ($\Gamma$) against Eddington ratio in our sample, overplotted on the Jin et al.\ (2012) sample. The red dashed and black long dashed lines are the best fit linear relations for our sample and the Jin et al.\ (2012) sample respectively, with the grey and pale red shaded regions being the 1 $\sigma$ error ranges of these relationships. The green solid line shows the relation derived by Shemmer et al.\ (2008) and Zhou et al.\ (2010).}
	\label{fig:xr_edd}
\end{figure}

We show a comparison of our models with those of \cite{jin12_1} for the \mdot--$\Gamma$ relationship in Fig. \ref{fig:xr_edd}. Due to the small size of our sample, there is a large uncertainty on the slope of this relationship, but it shows a correlation which is in agreement with that presented in \cite{shemmer08}, \cite{zhou10} and \cite{jin12_1}.

\nocite{shemmer08}
\nocite{zhou10}

\begin{figure}
	\centering
	\includegraphics[trim=0cm 2.5cm 0cm 3cm, clip=true, width = 0.5\textwidth]{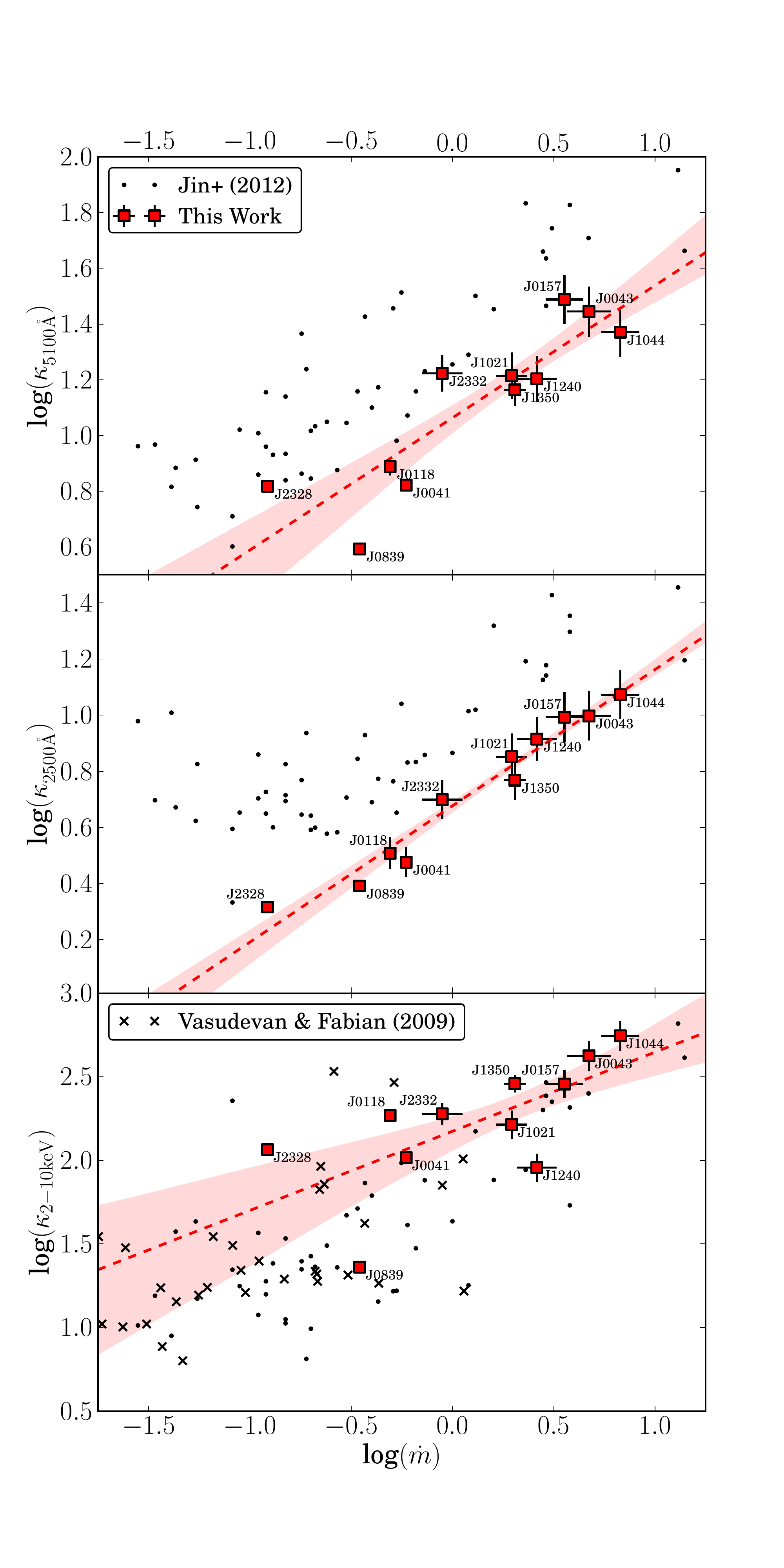}
	\caption{\small Dependence of various BC factors on \mdot. In addition to our sample, we show the literature samples of Jin et al.\ (2012) and Vasudevan \& Fabian (2009). The best fit linear trend lines for our sample are shown by the red dashed lines and the shaded regions show the associated 1 $\sigma$ error ranges.}
	\label{fig:kbols}
\end{figure}

We see a large spread in the BCs in our sample (Table \ref{tab:sedmod}), as previously mentioned in Paper I. Early work used fixed values for these coefficients (\eg \citealt{elvis94}, \citealt{richards06}), but cautioned that the large dispersion in these values made their application uncertain. There is mounting evidence that BCs are dependent on AGN luminosity, and by extension \mdot (\citealt{trakhtenbrot12}). Moreover the \cite{elvis94} SED templates included the IR torus bump in \Lbol, and therefore `double-counted' some of the emission from the AGN (\citealt{marconi04}). We present our BCs against \mdot in Fig.\ \ref{fig:kbols}, confirming the correlations observed by \eg  \cite{vasudevan09}, \cite{jin12_3} and \cite{castello-mor16}. We have overplotted the results of \cite{vasudevan09} and \cite{jin12_3} for direct comparison.

\cite{jin12_3} also used \optxagnf, applied to a low redshift ($z<0.3$) sample, whereas \cite{vasudevan09} used a simpler AD$+$PLT model. As our samples were all selected by different means, there may be differing selection effects between the samples. For instance, we required objects that were bright enough to yield an X-ray spectrum, \cite{jin12_3} imposed X-ray quality cuts (sample selection described in \citealt{jin12_1}), and \cite{vasudevan09} drew their sample from the \cite{peterson04} sample -- an RM study of optically bright AGN.

In the top two panels of Fig.\ \ref{fig:kbols}, our BC values for $\kappa_{5100 \rm \mathring{A}}$ and $\kappa_{2500 \rm \mathring{A}}$ show strong correlation with \mdot. This alone suggests that \Lbol can be constrained with an estimate of \M_BH and a measurement of the (dereddened) continuum luminosity. However, our values for both coefficients lie below the majority of the \cite{jin12_3} sample. As these are calculated from luminosity measurements that are AD dominated, the likely reason for this is the different average \M_BH of our two samples. The \cite{jin12_3} sample contains AGN with a lower average \M_BH; notably, many of the highest \mdot AGN in their sample were the narrow-line Seyfert 1 galaxies, with masses of $10^6-10^8 M_{\odot}$. The highest \mdot AGN in our sample are $\sim 1-2$ dex more massive, so a single AD continuum luminosity measurement samples a different part of the AD SED in our AGN, compared to theirs. Our BCs sample continuum regions closer to the AD peak, and are therefore smaller on average than those calculated by \cite{jin12_3}. This is illustrated in \cite{davis11}, their Fig.\ 1.

In the bottom panel of Fig.\ \ref{fig:kbols}, our results for $\kappa_{2-10 \rm keV}$ are more consistent with those of \cite{jin12_3} and \cite{vasudevan09}, suggesting that this coefficient depends less on effects such as \M_BH, as might be expected from the argument above -- \M_BH does not influence the X-ray spectrum as much as the AD.

In summary, we suggest that UV BCs in AGN are dependent on both \M_BH and \mdot, and applying relations calibrated for $\sim 10^7 M_{\odot}$ AGN to those with $\sim 10^9 M_{\odot}$ BHs could introduce systematic uncertainties. This may be complicated further by model-dependencies such as \astar, since $L_{\rm bol}=\eta \dot{M} c^2$, where the mass--energy efficiency, $\eta$, varies with \astar. X-ray BCs do not appear susceptible to this systematic effect, but do show a larger spread.

We will further explore the correlations between SED properties and BCs in a future paper, with a much larger AGN sample.

\subsection{SED Model Testing} \label{subsec:sed_test}

\subsubsection{Intrinsic reddening} \label{subsub:red}

Our SED model depends on the adopted models for intrinsic reddening in the AGN, and in Section \ref{subsec:intred} we showed that MW, LMC and SMC reddening curves are adequate to model the intrinsic extinction in all 11 of our AGN.

An alternative approach is to calculate customised extinction curves. \cite{zafar15} carried out a study of the intrinsic reddening of 16 quasars in the redshift range $0.71<z<2.13$, selected on the basis of high intrinsic extinction. By comparing their sample of objects to the \cite{vandenberk01} and \cite{glikman06} quasar templates, they were able to derive reddening curves for each object in their sample. However, an assumption in that work is that the intrinsic SED of all AGN in the sample can be described by a simple power-law of constant slope. Whilst a power law well-describes the optical--NIR continuum emission for many AGN, the true continuum is more accurately described by the AD, which has a predicted turnover in energy corresponding to the temperature of gas orbiting the BH just outside \risco , which is dependent on \M_BH, mass accretion rate and spin (\eg \citealt{hubeny00}, \citealt{davis07}). In Paper I we found that around half of the objects in our sample had optical spectra at or very near this SED peak. Additionally, we have found evidence for a change in power-law slope in 8 of 11 objects, consistent with the observations sampling the outer edge of the AD. For these reasons, we cannot make assumptions about the intrinsic SED shape {\it a priori}. As the intrinsic extinction in each of our objects is small -- \mbox{$E$($B-V$)$_{\rm int}$ $<0.1$ mag} in all but one case -- we are justified in our approach of using MW, LMC and SMC curves.

By allowing the dust composition to vary, we have made a logical extension to the modelling used in Paper I. Whilst some objects (such as J1350$+$2652) show evidence for a 2200 \AA \ feature that is better fit by a MW reddening curve (Paper I; a similar finding is shown in \citealt{capellupo15}, their Fig.\ 7), J1044$+$2128 lacks this feature and shows a much improved extinction correction with the SMC curve.

\subsubsection{Uncertainties on the black hole mass} \label{subsub:mass}

We have shown that in models of this kind, \M_BH uncertainties of $\sim 0.1$ dex lead to a $\sim 0.1$ dex uncertainty in \Lbol. In objects with well-sampled SED peaks, the difference is much smaller, as other parameters are adjusted to maintain the fit. This is perhaps clearest in J0839$+$5754, but also J0041$-$0947 and J0118$-$0052 show only small changes in \Lbol, despite a $\sim 0.4$ dex change in \M_BH from largest to smallest values. However, we should note that in J2328$+$1500  the optimal solution shows some degeneracy between \mdot and the intrinsic reddening as defined by $E$($B-V$)$_{\rm int}$, with the latter property converging to different minima when fitting the $\pm 2 \sigma$ \M_BH models. This may be expected in this object, as it shows the highest intrinsic reddening value (see Table \ref{tab:sedpar}) of our sample and therefore we suggest that for reddening of $E$($B-V$)$_{\rm int}$ $\gtrsim 0.1$ mag, the uncertainties in estimates of \Lbol become greater (a combined error due to \M_BH and $E$($B-V$)$_{\rm int}$ of $\sim 0.3$ dex in this case).

Our \M_BH estimates were specifically derived from the profile of \Ha as there is excellent signal-to-noise (S/N), and it shows strong correlation with \Hb, and hence reverberation studies \citep{greene05}. Assuming that the two main sources of uncertainty on \M_BH are the dispersion on the relation with \Hb and our measurement error may be optimistic (\citealt{park12} estimates that the uncertainty in the BLR size-luminosity relation and virial coefficient contribute to a total uncertainty on such estimates of $\sim 0.46$ dex), but we wished to test how such uncertainties would affect the calculation of the SED model, and corresponding properties. \cite{castello-mor16} used RM \M_BH estimates for their samples of super- and sub-Eddington AGN, but estimated the uncertainty on these estimates was still a factor of $\sim 3$, comparable to the single-epoch method.

\begin{figure}
	\centering
	\includegraphics[trim=0.6cm 0.0cm 0.7cm 1.0cm, clip=true, width = 0.5\textwidth]{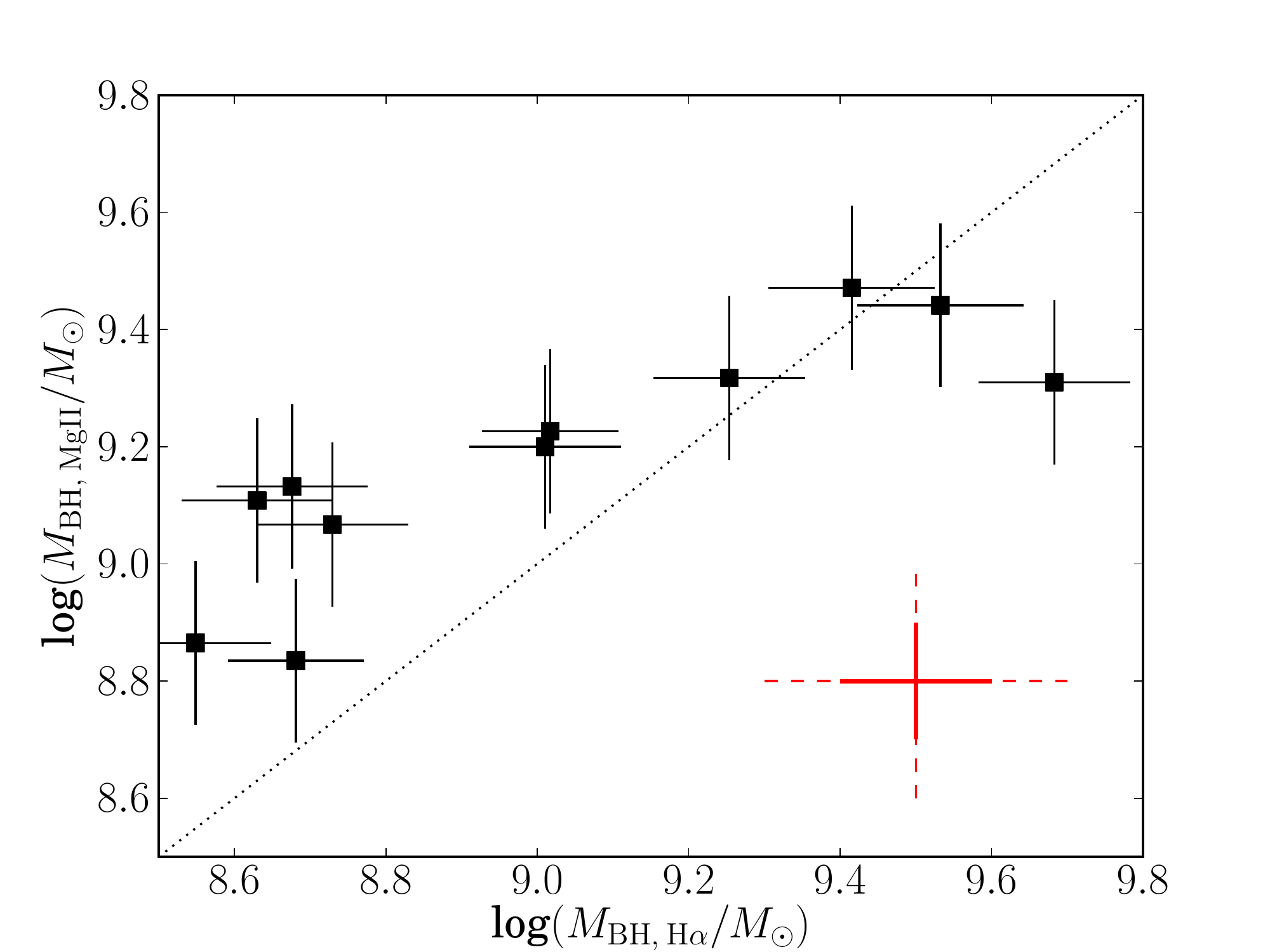}
	\caption{\small Comparison of mass estimates determined in Paper I from \Ha emission line with those using \Mgii measurements from Section \ref{sec:spec_decomp}, and the method of McLure \& Dunlop (2004). Error bars are representative of the intrinsic errors from the dispersions of the emission line relations, tabulated in Table \ref{tab:sedmod} for \Ha and taken to be 0.14 dex for \Mgii (Wang et al.\ 2009). The total uncertainties, when factoring in the error on RM mass measurements against which such estimates are calibrated, is larger. In red we show representative 1 $\sigma$ (bold) and 2 $\sigma$ (dashed) errors considered in Section \ref{subsec:mass}. There is some evidence for \Mgii masses being systematically larger at the low end of the scale, which could be statistical, or a feature of the specific relations we have used.}
	\label{fig:mgii_mass}
\end{figure}

We further explored the potential uncertainties on \M_BH by calculating new mass estimates using the \Mgii emission line and the method of \cite{mclure04}. \Mgii properties were determined in Section \ref{sec:spec_decomp}, and the results are shown in Fig.\ \ref{fig:mgii_mass}. The intrinsic dispersion on the \Mgii mass estimate was taken to be 0.14 dex (\citealt{wang09}). From this comparison we suggest that the $\pm 1,\:2\: \sigma$ uncertainties we considered in Section \ref{subsec:mass} are a reasonable representation of the error on \M_BH. There is some evidence for a trend, with \Mgii masses being systematically larger in the lowest mass objects of our sample. This could be statistical (the error bars plotted in Fig.\ \ref{fig:mgii_mass} are likely smaller than the true intrinsic uncertainties), or it could be due to the specific relations we use to determine \M_BH, which can be sensitive to the manner of the spectral analysis -- see \cite{wang09}. Using the relations of \cite{shen11} or \cite{trakhtenbrot12} may accentuate this effect even further, as those studies present even greater \M_BH values from \Mgii than \cite{mclure04}.

Finally we note that the profile of \Ha in J1044$+$2128 (and to a smaller degree J0043$+$0114 and J1021$+$1315, see Figs.\ \ref{fig:specdecomp1} and \ref{fig:specdecomp2}) suggests that addition of another Gaussian component may result in an improved fit. If such a component is associated with a `narrow' \Ha region (distinct from the BLR), this results in a broader `broad' component, and hence a ($\sim 0.1$ dex) larger \M_BH. However, given the extremely weak \Oiii line, it is not certain whether such a third \Ha component should indeed be physically attributed to a separate region; our data are insufficient to unambiguously determine its velocity width.

\subsubsection{Black hole spin} \label{subsub:spin}

Increasing the BH spin has a similar effect to lowering \M_BH  -- both increase the peak temperature of the AD gas, extending the SED peak to higher frequencies. However, spin also changes the efficiency, so that the same mass accretion rate through the outer disc (sampled by the rest-frame optical/UV spectra) will produce a higher \Lbol, as the disc extends closer to the BH. In our model, this affects the SX and PLT as well as the AD, since these are assumed to be powered directly by the same accretion flow observed as a disc in the optical/UV. This is done via the \rcor parameter, which sets the radius below which the luminosity is used to power these soft and hard X-ray components, rather than being dissipated in a standard AD. Thus the \optxagnf model has a peak disc temperature which is set by \rcor, rather than by BH spin directly. Increasing the spin means that there is more energy dissipated below \rcor, \ie there is more energy to power the SX and PLT. Since the level of the PLT is fixed by the X-ray data, this means that the fit generally adjusts \rcor to smaller values, leading to an increase in peak disc temperature compared to zero spin. This makes the (rest-frame) UV spectrum bluer, so the intrinsic reddening decreases to maintain the fit.

For low spin values ($a_* \leqslant 0.5$), these adjustments are minor and the fits are similar to those with zero spin. However, for higher spins ($a_* = 0.8$ and $0.9$), this has a large impact on the models, with the bluer UV continuum being very different to the observed continuum slope in a way which cannot be easily compensated for by decreased reddening. In half of the sample, this produces a poorer fit to the data, but this is not true for all objects; J0041$-$0947, J0043$+$0114, J1240$+$4740, J1350$+$2652, J2328$+$1500 and J2332$+$0000 all show reasonable fits (\rchi$<3$) for the $a_* = 0.9$ model. The resulting fits are markedly poorer for the highest spin states ($a_* \geqslant 0.99$), ruling these out from the \optxagnf modelling.

A limitation in our study is that we have not considered the effect of AD inclination in our models. The \optxagnf model assumes a constant inclination, $\theta$, of $60^{\circ}$ to face-on, and geometric consideration of orientation dictates that a factor of two greater flux would be observed if the AD was face-on ($\theta = 0^{\circ}$). Larger inclinations than 60$^{\circ}$ are thought to be less likely, as at some point the coaxial torus would obscure the AD. The effect of this on the SED peak frequency will be small, making this property extremely difficult to robustly constrain and practically, other sources of uncertainty discussed in this chapter dominate the uncertainty on \Lbol.

The disc inclination becomes significant in the case of a highly spinning BH. Here, relativistic effects arising from the differential line-of-sight gas motion at different inclinations must be accounted for, as the simple trigonometric treatment of inclination is insufficient. It is therefore necessary to convolve a relativistic smearing kernel with the AD spectrum at each radius. This formed the basis of the model presented in \cite{done13} that includes such relativistic treatment of the AD inclination  -- \mbox{\optxconv}.

\begin{figure}
	\centering
	\includegraphics[trim=0.3cm 0.0cm 0.5cm 1.1cm, clip=true, width = 0.5\textwidth]{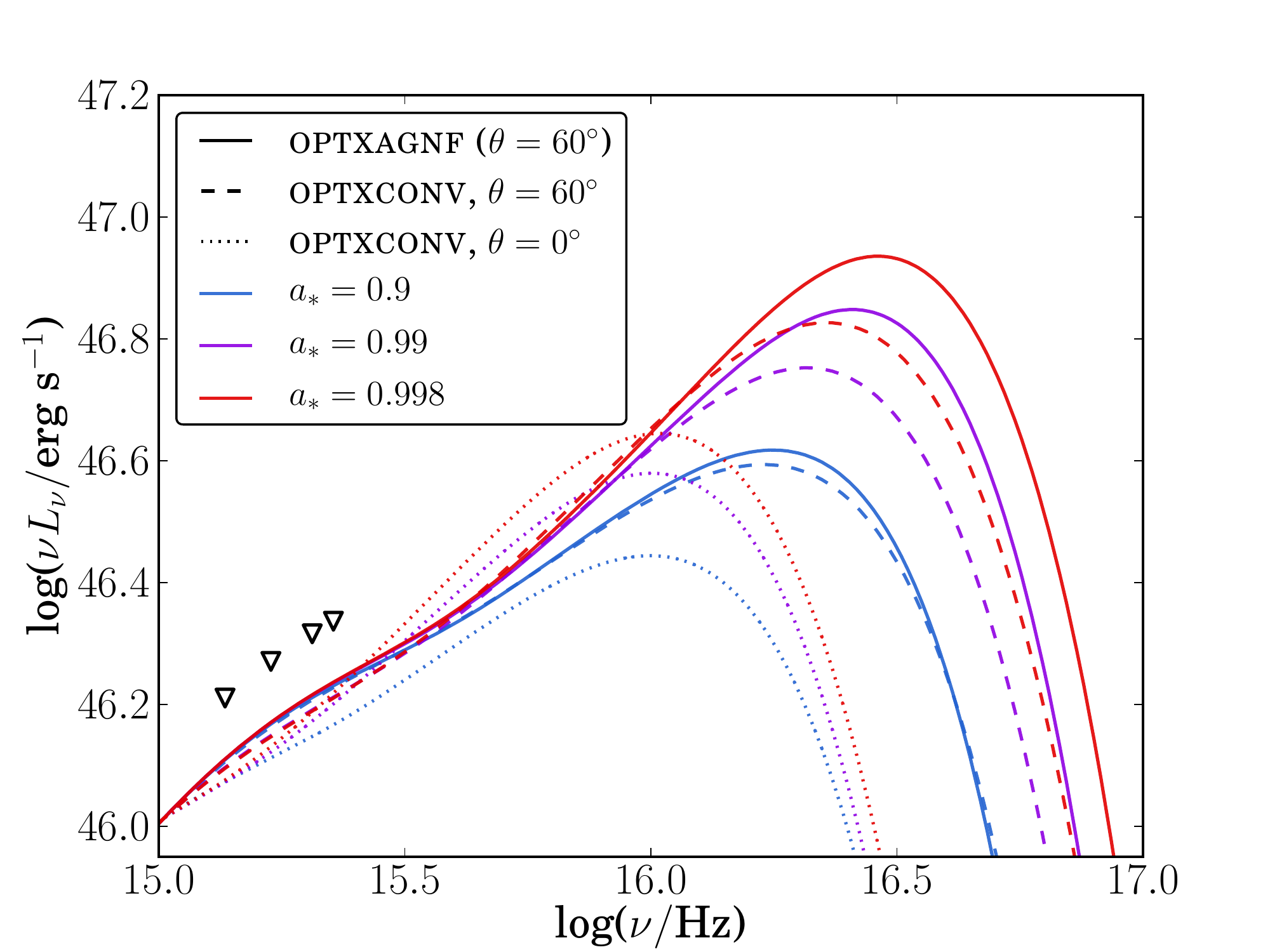}
	\caption{\small Comparison of \optxagnf with \optxconv for high to maximal spin BHs ($M_{\rm BH}=10^9 M_{\odot}$, $\dot{m}=1$). \optxconv includes relativistic effects that are neglected in \optxagnf. The line styles (solid, dashed, dotted) denote the model used, and the colour corresponds to different values of \astar. To more clearly show the difference in the extreme-UV, the models are arbitrarily normalised to the same value at $10^{15}$ Hz, as this part of the SED is constrained by data. The markers show the wavelengths used for SED fitting, highlighting the difficulty of distinguishing between these scenarios with the available data.}
	\label{fig:agnf_vs_conv}
\end{figure}

In Fig.\ \ref{fig:agnf_vs_conv} we compare models of \optxagnf (no relativistic convolution) with \optxconv (which includes the relativistic convolution) at high to maximal spin values. We normalise all models at a frequency of $10^{15}$ Hz (3000 \AA), as data constrains this part of the model, and they must all therefore pass through the same point. We show two different inclinations for \optxconv, $\theta = 0^{\circ}$ and $60^{\circ}$, as the difference between the two models is strongest in the case of a face-on disc. Due to this discrepancy in energy, it is possible that more of our objects could be compatible with high spin ($a_* \geqslant 0.99$) values than we concluded in Section \ref{subsec:spin}.

Therefore, we reproduced the high spin models described in Section \ref{subsec:spin} using \optxconv at inclinations of both $0^{\circ}$ and $60^{\circ}$. We tested spin values of $a_* = 0.99$ and $0.998$, as both scenarios were ruled out in most objects when using \optxagnf. We found that \optxconv delivers similar models to \optxagnf when the inclination was fixed at $60^{\circ}$; either a good fit to the data is not achieved, or the SED energy is dominated by the SX and PLT, in spite of the high accretion rates. This is counter to what is observed at lower redshifts (\citealt{jin12_1}), in which high accretion rate objects are generally AD dominated. In the face-on case, however, good fits to the data are obtained in 10/11 AGN (J1044$+$2128 is the exception), even for $a_* = 0.998$. Four of these AGN still require a dominant SX and PLT, however. We therefore highlight that models including full relativistic treatment of the disc inclination should be used to model highly-spinning BHs, as the difference in energy can be significant. However, as we have already noted some sources of potential degeneracy between parameters in our models, it is unlikely that a strong constraint can be put on the AD inclination by SED modelling alone. Using this method to accurately measure the accretion rate, spin, inclination and intrinsic reddening values would require exceptional data coverage and quality. Despite these uncertainties, our measurements of \Lbol are more accurate than those in other studies that lack X-ray spectra, in addition to optical--NIR spectra (\citealt{capellupo15}, \citealt{castello-mor16}).

So, do our results support a `spin-up' picture of BH evolution? If BHs grow via prolonged anisotropic accretion episodes and mergers with other BHs, their spin values would be expected to increase over cosmic time, such that the most massive BHs have the highest spins (\eg \citealt{dotti13}, \citealt{volonteri13}). The counter argument is that randomly oriented accretion episodes would result in $a_*$ approaching zero for massive AGN (\eg \citealt{king08}). An alternative finding by \cite{fanidakis11} suggests that prolonged accretion episodes spin up \textit{all} supermassive BHs, whilst chaotic accretion results in only the most massive ($\gtrsim 10^8$ \Msun) BHs having high spins, as a result of merger-driven growth.

Using the results from \optxagnf, we find our AGN to be more consistent with having low to moderate spins. However, high spins cannot be ruled out for face-on inclinations when relativistic corrections are included in \optxconv. We can still conclude that the most massive AGN in our sample are all compatible with hosting highly spinning BHs, whereas the least massive (J1044$+$2128) is not, but reiterate that there are several sources of degeneracy in the models. 

\subsubsection{Radial extent of accretion disc} \label{subsub:rout}

We find that in the eight AGN where we put constraints on \rout with our model, there is a strong correlation with \rsg, but offset from unity, see Fig.\ \ref{fig:cf_rout}. It is not known whether self-gravity is the condition under which the disc breaks up, but our findings suggest that \rout could be related to \rsg, but smaller by a factor $\sim 5$, in most or all cases.

This result differs from that in \cite{hao10}, who study the optical-IR SEDs of a sample of `hot-dust-poor' AGN. In a quarter of their sample, weak host galaxy contribution enables measurement of the outer AD radii, which are found to be larger than \rsg. This could suggest a difference in the AD in these objects that may or may not be related to their weak dust contributions. Alternatively, it could be a result of poorer data coverage, as they use photometry for their SED fits, or degeneracy with dust blackbody components. We require a greater understanding of AD physics to unify these observables.

For this test, we kept the spin fixed at zero, but \rsg is determined by the AD total mass, which is not very dependent on BH spin. It does however depend on the assumed Shakura-Sunyaev viscosity parameter, $\alpha$. We assume $\alpha=0.1$; a smaller value of $\alpha$ would result in a more massive disc, and hence smaller \rsg. However, this dependence is not very strong, and requires $\alpha \sim 10^{-4}$ to account for the difference we infer. 

\subsection{Torus and Host}

\begin{figure*}
	\centering
	\includegraphics[trim=3.3cm 0cm 3.3cm 0cm, clip=true, width = \textwidth]{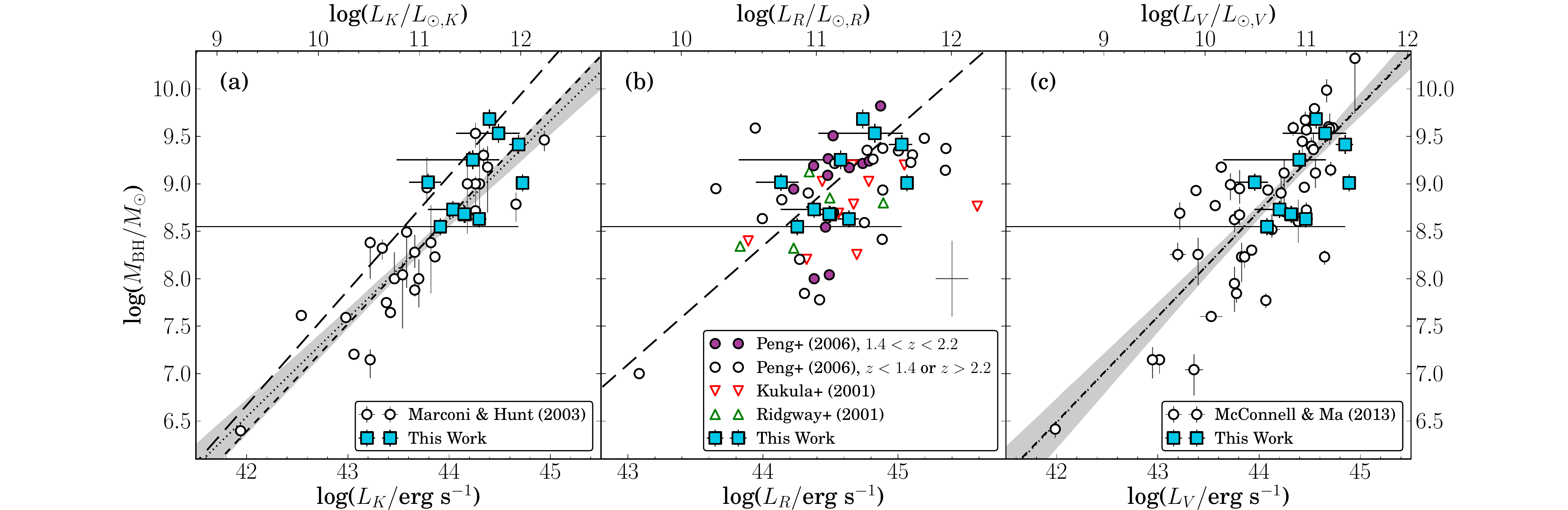}
	\caption{\small Comparison of modelled host galaxy luminosities with the Marconi \& Hunt (2003), Peng et al.\ (2006) and McConnell \& Ma (2013) samples (panels (a), (b) and (c) respectively). In (b), we show a representative error bar for the literature data, and highlight the objects in the same redshift range as our own in purple. Also shown in (b) are the literature samples of Kukula et al.\ (2001) and Ridgway et al.\ (2001). In panels (a) and (c) we show the derived relations by each work (short dashed lines) together with a simple least squares regression line (dotted line) with 1$\sigma$ error, estimated by drawing 1000 bootstraps from the literature sample. In panels (a) and (b) we show for comparison the relations derived respectively by Kormendy and Ho (2013) and Peng et al.\ (2006) for local AGN (long dashed lines). Peng et al.\ (2006) used literature data from Kormendy \& Gebhardt (2001) and Bettoni et al.\ (2003) for this calculation.}
	\label{fig:cf_host}
\end{figure*}
\nocite{kukula01}
\nocite{ridgway01}
\nocite{kormendy13}

The mean temperatures for the two blackbody components that model the torus for our sample are $\langle T_{\rm hot} \rangle = 1470\pm90$ K and $\langle T_{\rm warm} \rangle = 430\pm30$ K. The warm component is generally poorly constrained by only two \textit{WISE} photometry points, and shows large errors on $T_{\rm warm}$. Our mean covering factors are $\langle C_{\rm hot} \rangle = 11\pm2$ per cent and $\langle C_{\rm warm} \rangle = 14\pm3$ per cent.

\cite{landt11} obtain values of $\langle T_{\rm hot} \rangle = 1365\pm18$ and $\langle C_{\rm hot} \rangle = 7\pm2$ per cent. The \cite{landt11} sample had NIR spectra from the NASA Infrared Telescope Facility's SpeX spectrograph, and as such the data was of significantly higher quality than was available for this work, for which the torus components were only constrained by \textit{WISE} photometry. Their sample was also at a lower redshift ($z \leqslant 0.3$) and lower average luminosity than our sample. Nevertheless, our results are consistent to within $2 \sigma$.

As discussed in \eg \cite{landt11}, \cite{burtscher13}, \cite{kishimoto13} and \cite{netzer15}, the $T_{\rm hot}$ values calculated are close to the silicate dust grain sublimation temperature. This may suggest the grains were formed in an oxygen-rich environment. We do see evidence for a spread in $T_{\rm hot}$ values which is likely to be a feature of the limited quality of our data, although \cite{landt11} note that in NGC 5548 there is some evidence for higher dust temperatures than other objects in their sample. Other studies finding similar results for $T_{\rm hot}$ include \cite{kobayashi93} (using a similar approach to ours), and \cite{suganuma06}.

\cite{mor11} found their distribution of hot dust covering factor values peaked at $\sim 13$ per cent, in a sample of 15,000 SDSS AGN, fitting \textit{WISE} photometry of comparable quality to our sample. This is slightly higher than the \cite{landt11} value, but is consistent with our result, which lies between the \cite{landt11} and \cite{mor11} values. The \cite{mor11} sample covers a larger range of luminosities than ours but they do not find a dependence of covering factor on \M_BH or \mdot.

Finally, \cite{roseboom13} inferred a broad distribution of covering factors, generally greater than those measured in this work, \cite{landt11} and \cite{mor11}. They fit the AGN component using \cite{elvis94} SED templates, rather than the physically motivated model we employ for our analysis, and this may lead them to underestimate the AGN luminosity, and predict higher covering factors. Studies of the covering factor are important in the context of exploring the receding torus scenario proposed by \cite{lawrence91}, where the covering factor is dependent on the AGN luminosity. While there is currently little evidence for this (\citealt{mor11}, \citealt{netzer15}), approaches such as ours provide a means of testing this with pre-existing data.

To assess whether the host galaxy properties we predict are reasonably concurrent with other research of AGN and their hosts, we compare the galaxy luminosities we have calculated with earlier work. From our fitted host galaxy model, we measure the luminosity in the $V$, $R$ and $K$ bands, comparing these to the results presented in \cite{marconi03}, \cite{peng06} and \cite{mcconnell13}. We measure host galaxy luminosities using a `synthetic photometry' technique, by integrating the fitted template over the respective bandpass. Our host galaxy luminosities and \M_BH estimates are shown, together with those of the literature samples in Fig.\ \ref{fig:cf_host}.

Our data are broadly in agreement with the \M_BH--bulge relationship. The \cite{peng06} sample is of particular interest, as several objects in their sample are at comparable redshifts to our study. In Fig. \ref{fig:cf_host} (b), the dashed regression line we show is derived by \cite{peng06} from a sample of 20 nearby AGN, with \M_BH and $L_R$ values published in \cite{kormendy01} and \cite{bettoni03} respectively. The \M_BH values calculated by \cite{peng06} are made using the single epoch virial linewidth technique, however, due to lack of IR spectra, \M_BH is estimated from the emission line profiles of \Civ and \Mgii in several objects. Where both of these lines were available, \cite{peng06} used the average \M_BH estimate from these lines, whereas in Fig. \ref{fig:cf_host} we show the \Mgii estimate only, as \Civ has been consistently shown not to correlate well with the \M_BH estimate from the Balmer lines (discussed more in Section \ref{subsec:spec_decomp}). Additionally, some of the \cite{peng06} \M_BH estimates were made manually by the authors of that study from printed copies of the spectra, and as such these may be less reproducible than the Gaussian fitting method we employ. In spite of these uncertainties, our two samples from this redshift range are both consistent with the low redshift relationship.

\cite{paltani98} and \cite{soldi08} note that in the AGN 3C 273, variability suggests there are two distinct contributions to the optical-UV continuum. Whilst the consequences of this finding are unclear, it could suggest an additional contribution to the SED between the AD and torus. In our study, it is probable we would end up attributing such a contribution to the host galaxy. Once again, our data are not sufficient to support or contradict such a result, although if the AD does truncate at the relatively small radii measured in Section \ref{subsec:rout}, this could provide additional matter to form such an additional component.

\begin{figure*}
	\centering
	\includegraphics[trim=2cm 0cm 2cm 0cm, clip=true, width = \textwidth]{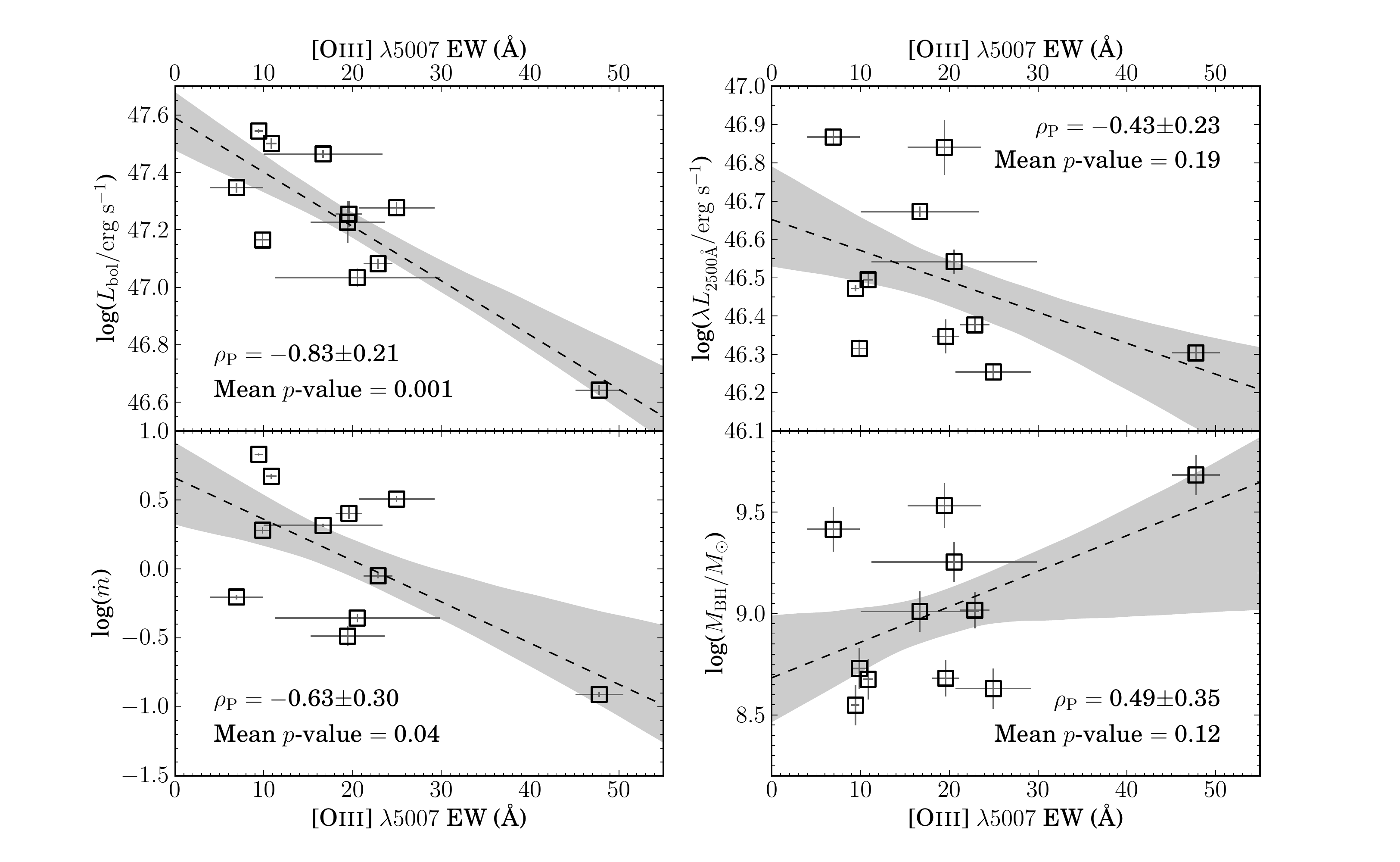}
	\caption{\small Comparison of \Oiii $\lambda 5007$ EW with various AGN and SED properties. Best-fit regression lines and error ranges determined from our bootstrap method are shown. A statistically significant anti-correlation is seen with the bolometric AGN luminosity, \Lbol. Based on our data, this is the property most strongly linked to \Oiii line strength. The other properties show weaker relations and, to $\sim 2 \sigma$, are consistent with being uncorrelated.}
	\label{fig:oiii_ew}
\end{figure*}

In the IR, only the \textit{WISE} photometry and part of the NIR spectrum confines the torus and host galaxy components, but this has still proven adequate to set useful constraints on these. This opens up a range of possibilities for examining correlations between the central engine, torus and wider galaxy in larger AGN samples. Our technique could be applied to many AGN with NIR and optical spectra, and \textit{WISE} photometry, and greatly expand investigations such as \cite{peng06} at higher redshifts, as it does not require \hst imaging of gravitationally lensed galaxies.

We have only adopted the 5 Gyr elliptical galaxy template from \cite{polletta07}. Our assumption that this is a plausible host galaxy class is based on local scaling relations that may not hold at the redshift range of our sample, and a future study may incorporate alternative templates to probe these relations further. However, in Paper I we found that the practical differences between starburst and elliptical galaxy templates for our data quality were small, with the additional UV emission related to star formation contributing $\sim 1$ per cent of the AGN flux at $\sim 2000 \; \rm \mathring{A}$ rest frame, and so the analysis presented here ought to be sufficient. There are now refined templates available, such as those presented in \cite{brown14a}. Once again, given our data quality and the dominance of the AGN/torus in this region, the differences between host galaxy templates are not significant for our purposes.

\subsection{Spectral decomposition} \label{subsec:spec_decomp}

Using our SED continua, we have undertaken a spectral decomposition of the optical--NIR data for our 11 objects.

Firstly, it is clear that around half of our objects have weak narrow \Oiii, which appears at first to be anti-correlated with the Eddington ratio. In general, the lowest accretion rate objects show the strongest narrow emission lines (J0839$+$5754 and J2328$+$1500 are the clearest examples). Similarly, the highest accretion rate objects (particularly J0043$+$0114, J1021$+$1315, J1044$+$2128 and J1240$+$4740) show extremely weak narrow \Oiii. (The narrow feature at $\sim5000$ \AA \ in J1021$+$1315 is attributed to noise, as there is a corresponding feature in the error array.)

We explored this further by searching for (anti-)correlations between the \Oiii $\lambda$5007 line equivalent width (EW) and \mdot, \Lbol, \M_BH and $L_{\rm 2500 \mathring{A}}$. We show plots of these properties in Fig.\ \ref{fig:oiii_ew}, and use the Pearson product moment correlation coefficient, $\rho_{\rm P}$, and $p$-value to assess whether correlations are statistically significant. To estimate the uncertainty of the relations, we draw 2000 bootstrap subsamples, repeating the analysis on each of these, and taking the central 68 per cent of the resulting distributions as an indication of the 1 $\sigma$ error on each property. Using the deviance of $\rho_{\rm P}$ from zero as an indicator of (anti-)correlation between properties, we see that the strongest anti-correlation is between \Oiii EW and \Lbol, at almost 4 $\sigma$ significance. (Anti-)correlations between \Oiii EW and $L_{\rm 2500 \mathring{A}}$, \mdot and \M_BH are more uncertain, and appear to be largely dependent on a single object (J2328$+$1500). To within $\sim 2 \sigma$, these relations are consistent with no correlation.

\begin{figure*}
	\centering
	\includegraphics[trim=2cm 0cm 2cm 0cm, clip=true, width = \textwidth]{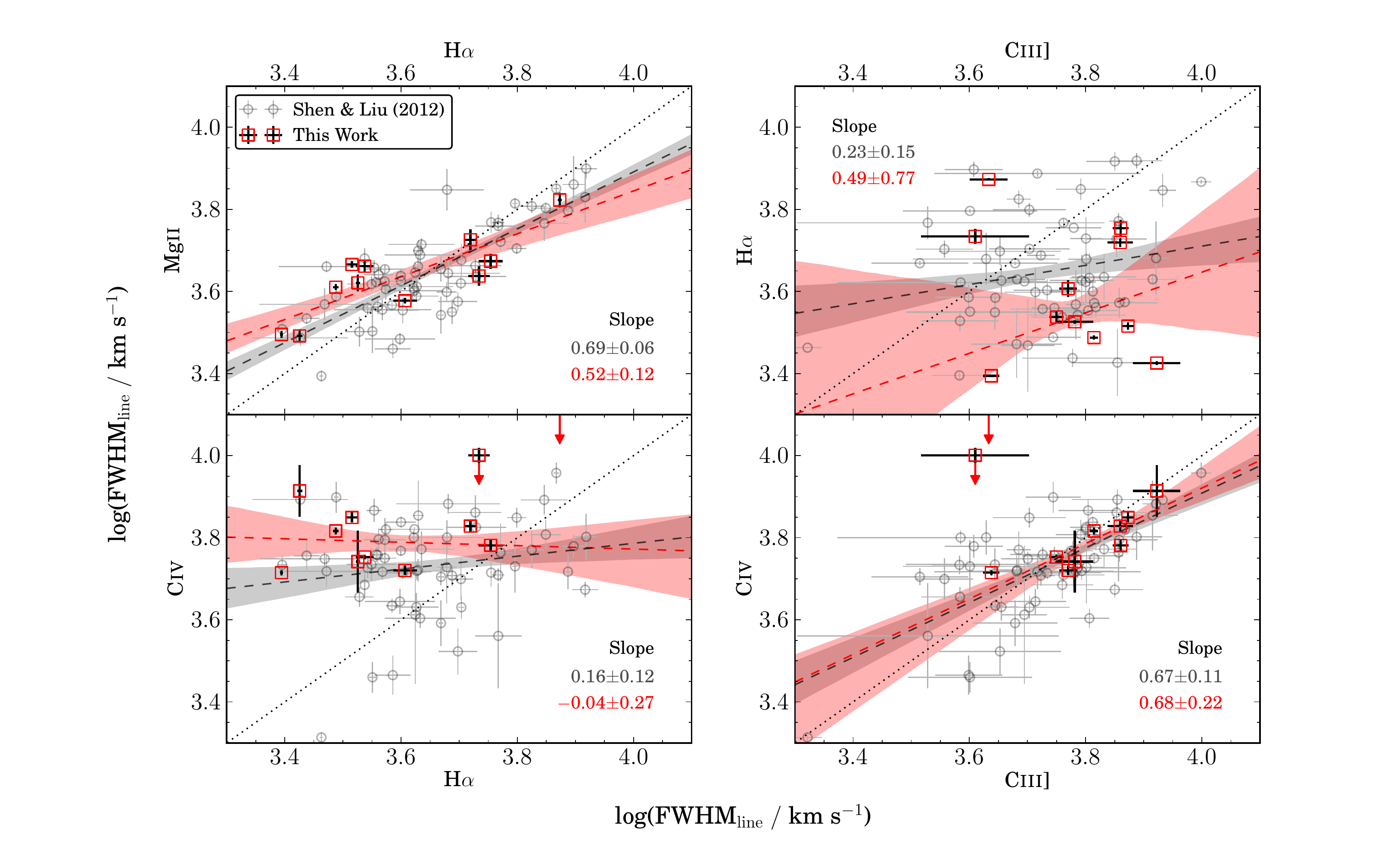}
	\caption{\small Comparison of calculated FWHM with the Shen \& Liu (2012) sample. Although four objects are common to both samples, we wish to test whether the two different means of characterising the AGN continuum (SED model versus power-law) are consistent.}
	\label{fig:cf_fwhm}
\end{figure*}

This may suggest that a narrow line region (NLR) in the most luminous sources cannot form, due to radiation pressure from the AGN. The objects with the weakest narrow \Oiii lines are similar to the broad absorption line quasi-stellar objects in the \cite{boroson92b} sample. \cite{netzer04} studied the disappearing NLR in a sample of $2 \lesssim z \lesssim 3$ quasars with higher average luminosities than ours. They suggested that some of the most luminous sources lose their dynamically unbound NLRs, although in others star formation at the centre of the galaxy may produce a NLR with different properties to lower luminosity AGN. \cite{netzer04} defines objects in their sample with \Oiii $\lambda5007$ equivalent width of $\sim 10-80$ \AA \ as showing `strong' \Oiii, corresponding to $\sim 2/3$ of their sample. Adopting the same definition yields $8/11$ objects in our sample -- a comparable fraction.

We next test whether results from our approach are consistent with larger studies, which have extensively studied relations between various linewidths (a probe of velocity dispersion) and luminosities (a probe of BLR size). We specifically consider the \cite{shen12} results; although four of our objects are common to their sample, here we compare the different means by which we deconvolve the spectra.

A comparison of the FWHM of various emission lines for our sample are shown in Fig.\ \ref{fig:cf_fwhm}. We also show the results of \cite{shen12}. In this work, we consider FWHM rather than other proxies for the linewidth, such as the line velocity dispersion (the second moment of the emission line profile, see \citealt{peterson04}). Although the dispersion is found to present a more unbiased proxy of the gas motion (see \citealt{collin06} and \citealt{denney13} for a comparison of the two approaches), the necessity for high S/N spectra to accurately measure the line dispersion disfavour this against the FWHM (\eg \citealt{shen12}).

\begin{figure}
	\centering
	\includegraphics[trim=0cm 2.3cm 0cm 3.5cm, clip=true, width = 0.5\textwidth]{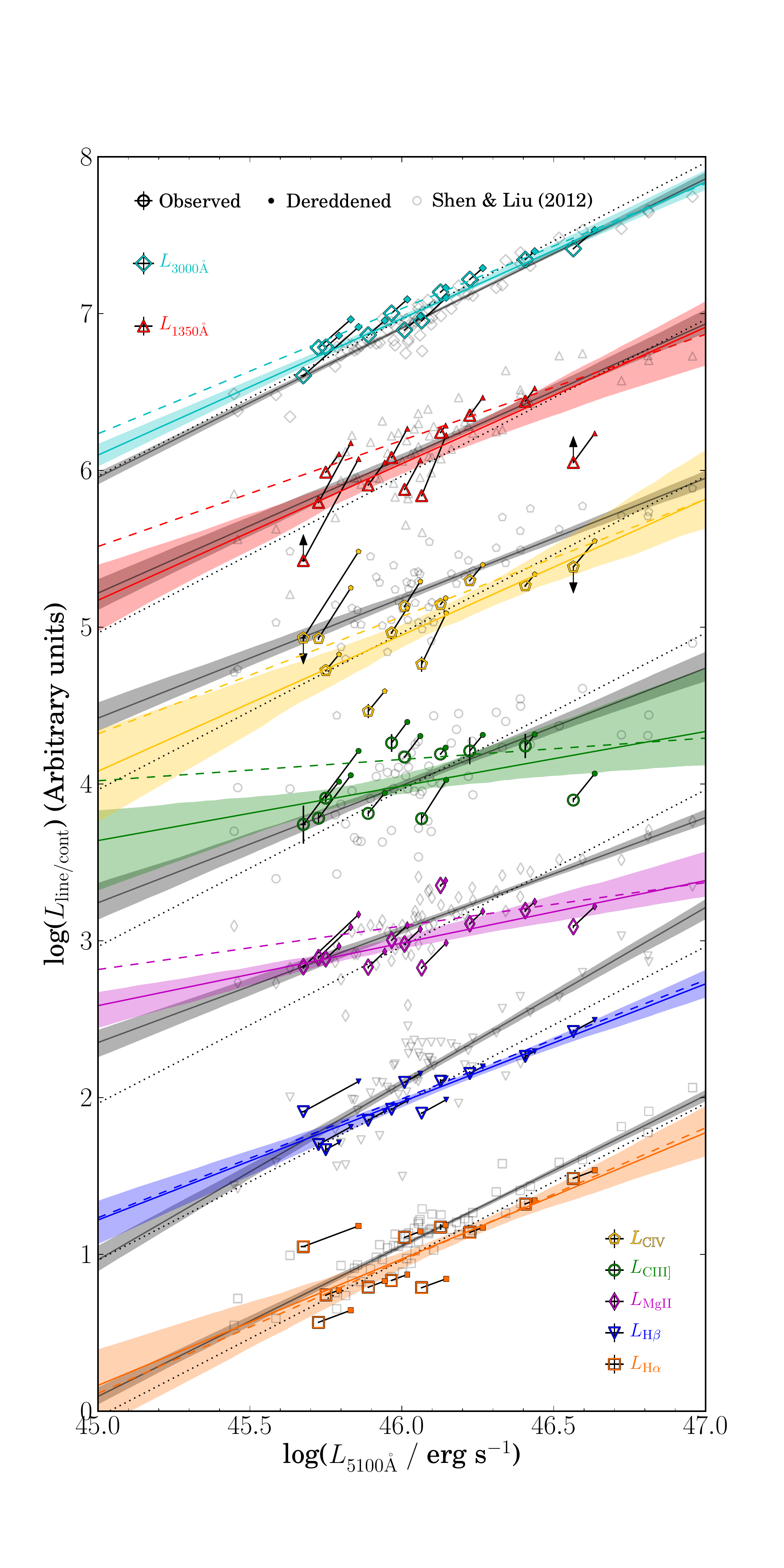}
	\caption{\small Comparison of luminosities with the Shen \& Liu (2012) sample. The observed data points are each linked to a corresponding point that shows the luminosity predicted after correcting for intrinsic extinction. Unity for each property is shown by the dotted lines. The Shen \& Liu (2012) sample is shown as grey symbols, with least squares regressions and $1 \sigma$ errors from 1000 bootstraps shown as the solid grey line and shaded region. Our data are shown in colour, with solid coloured line and shaded regions showing regression and $1 \sigma$ region. The dashed coloured lines are the dereddened regressions. Error bars are often very small, this is discussed in the text.}
	\label{fig:cf_lum}
\end{figure}

In Fig.\ \ref{fig:cf_fwhm}, we show least squares regression lines, together with $1 \sigma$ error region from drawing 1000 bootstrap subsamples from each distribution. Our sample regressions agree with the \cite{shen12} relations to within $2 \sigma$, demonstrating strong correlation between the \Ha and \Mgii FWHM, but no significant correlations with those for \Ciii or \Civ. There is also a correlation between \Ciii and \Civ FWHM. It has been previously noted that the \Civ line profile does not correlate well with \Hb (\eg \citealt{baskin05}, \citealt{netzer07}, \citealt{sulentic07}, \citealt{fine10}, \citealt{ho12}, but also see \citealt{vestergaard06}, \citealt{assef11}, \citealt{denney13} and \citealt{tilton13}), and \cite{shen12} also observe a correlation between \Ciii and \Civ. In two objects, J0839$+$5754 and J2328$+$1500, the full \Civ profile is not sampled by our data, and we measure very large \Civ linewidths. We therefore treat these results as upper limits, as we lack continuum measurements on either side of the emission line.

Our \Civ linewidths are systematically larger than the \Ha linewidths. This is expected from considerations of the BLR radius--luminosity relationship, but is not often seen (\citealt{trakhtenbrot12}).

The errors on our FWHM values are in general smaller than those determined by \cite{shen12}, even though both are calculated from similar Monte Carlo methods. This is likely due to \cite{shen12} using more components to model each line -- for \Civ, \Mgii and \Ha they use up to three Gaussians for the broad component (where we use only two) and one for the narrow component (which we do not model in \Civ or \Mgii, and only include in \Ha for objects with strong narrow \Oiii). This may lead to greater degeneracy between the components in their Monte Carlo fits, thus contributing to larger errors in FWHM.

In Fig.\ \ref{fig:cf_lum}, we show a comparison of our emission line and continuum luminosities against $L_{\rm 5100 \mathring{A}}$, and once again observe general agreement with the \cite{shen12} sample. We also show the predicted intrinsic luminosities after correcting for intrinsic extinction, connecting these points to the corresponding observed values with black lines. The least squares regressions through the observed points are shown with solid lines of corresponding colours, and the `dereddened' regressions are dashed lines. There appears to be no improvement in the relations for these dereddened values, and they appear in some cases to show poorer agreement with the unity line (dotted black lines). This reflects that the scatter introduced from considering the intrinsic extinction is larger than the scatter from adopting the luminosities as observed. In J0839$+$5754 and J2328$+$1500 we treat $L_{1350 \rm \mathring{A}}$ and \Civ luminosity measurements as limits, as they are not fully sampled by the SDSS spectra, and highly model dependent.

Once again, error values on our sample are very small. This is probably partly for the same reasons as discussed above -- fewer Gaussian components in the decomposition lead to less degeneracy -- but additionally our SED continuum contains only one free parameter (the normalisation), versus the \cite{shen12} approach, which uses power-law continua (in some cases with a break included), with both normalisation and slope left as free parameters.

\cite{mejia-restrepo16} used the \cite{capellupo15} data and models to directly compare results from performing a global decomposition with the AD model to those utilising local power-law continuum fits, finding the global approach to be more reliable. Although our study is smaller, we have also shown that using our physical model of the underlying AGN continuum facilitates a global spectral decomposition that is at least consistent with alternatives. Since we attach a physical significance to this component, compatible with accretion physics and constrained by multiwavelength data from mid-IR to X-rays, this is a better justified approach, compared to the empirical alternatives. 

\section{Summary and Conclusions} \label{sec:summary}

In this paper we substantially extend the study presented in Paper I with a systematic analysis of the SED parameter space. A physically motivated AGN SED model is applied to multiwavelength (mid-IR--X-ray) spectral data for 11 AGN in the redshift range $1.5 < z < 2.2$. A summary of our work and key findings is as follows:

\begin{itemize}
\item[i.]{We first refine the model described in Paper I to include best-fitting intrinsic extinction curves, out of MW, LMC and SMC models.}
\item[ii.]{The Eddington ratio -- photon index relation of our refined model agrees with previous work.}
\item[iii.]{We observe tight correlations between the UV bolometric correction coefficients and accretion rates. There is evidence for systematic offsets from the equivalent relation in a lower redshift sample, which we suggest is due to a higher average \M_BH in our sample. X-ray bolometric corrections are less susceptible to this effect, but show a larger spread.}
\item[iv.]{We next test the effect of uncertainties in the \M_BH estimate on the AGN bolometric luminosity (\Lbol). We find that in objects with well-sampled SED peaks, the difference is small, and in other objects, an uncertainty of $\sim 0.1$ dex in the \M_BH estimate propagates through to a $\sim 0.1$ dex uncertainty on \Lbol.}
\item[v.]{The effects of varying the BH spin parameter $a_*$ are explored. We find that spin values up to $a_*=0.9$ provide acceptable SED fits in 6 out of 11 objects (and an improvement over the $a_*=0$ model in 3), but that very high and maximal spin values of $a_* \geq 0.99$ are ruled out by a combination of the optical--NIR and X-ray data in all objects, with one exception. However, if we include relativistic treatment of the disc inclination, high and maximal spin values can describe the data in most objects, if the AD is face-on to the observer. There is degeneracy between the BH spin, inclination and mass accretion rate which make measurements of BH spin from continuum fitting uncertain.}
\item[vi.]{The outer disc radii are well constrained in 8 out of 11 objects. They show good correlation with the self-gravity radius, but are smaller by a factor $\sim 5$. This suggests that the disc break-up may occur closer to the BH than the self-gravity radius.}
\item[vii.]{We model the red end of the NIR to mid-IR ($2-22 \mu$m, observed frame) using host galaxy and torus models. We find good agreement with previous studies for both the torus properties (covering factor and temperature) and the host galaxy (luminosity). This is despite our more limited data-set in comparison with some of these investigations. We suggest that our approach to the SED modelling provides a viable alternative to structural decomposition of high-resolution images and those requiring observationally expensive mid-IR spectra.}
\item[viii.]{Our continuum model provides a firm basis on which to execute a spectral decomposition of the optical--NIR spectra. The results from our approach are in agreement with previous studies that utilise empirical models of the continuum. We see a statistically significant anti-correlation between \Oiii line strength and the AGN bolometric luminosity.}
\end{itemize}

\section*{Acknowledgements} \label{sec:acknowl}

The authors acknowledge the anonymous referee for a useful report that resulted in a greatly improved manuscript.

JSC is grateful to Kelly Denney and Allison Kirkpatrick for insightful conversations during the Sept 2015 AGN conference in Chania, Crete, and constructive comments from Beta Lusso. The authors would like to thank Marianne Vestergaard for kindly providing the UV \Feii templates from \cite{vestergaard01}, used in Section \ref{sec:spec_decomp}. JSC acknowledges the support of STFC grant ST/K501979/1. MJW and CD are supported by STFC grant ST/L00075X/1.

This publication makes use of observations and data products from the following facilities.

\begin{itemize}
\item The Gemini Observatory, which is operated by the Association of Universities for Research in Astronomy, Inc., under a cooperative agreement with the NSF on behalf of the Gemini partnership: the National Science Foundation (United States), the National Research Council (Canada), CONICYT (Chile), Ministerio de Ciencia, Tecnolog\'{i}a e Innovaci\'{o}n Productiva (Argentina), and Minist\'{e}rio da Ci\^{e}ncia, Tecnologia e Inova\c{c}\~{a}o (Brazil).
\item The Apache Point Observatory 3.5-metre telescope, which is owned and operated by the Astrophysical Research Consortium.
\item The \textit{Wide-field Infrared Survey Explorer}, which is a joint project of the University of California, Los Angeles, and the Jet Propulsion Laboratory/California Institute of Technology, funded by the National Aeronautics and Space Administration (NASA).
\item \xmmn, an ESA science mission with instruments and contributions directly funded by ESA Member States and NASA.
\item SDSS-III, funding for which has been provided by the Alfred P.\ Sloan Foundation, the Participating Institutions, the National Science Foundation, and the U.S.\ Department of Energy Office of Science. The SDSS-III web site is http://www.sdss3.org/.
\end{itemize}

Finally, we thank the contributors to the \python programming language and HEASARC, for software and services. This work made use of the online cosmology calculator described in \cite{wright06}.

\small{
\bibliography{2_SpecAn_v1}
\bibliographystyle{mn2e}
}

\appendix

\begin{table*}
	\section{AGN SED and Attenuation Model Parameters} \label{sec:app}
	\caption{\small Full table of parameters for the multi-component model described in Section \ref{subsec:intred}. The means of calculating the fixed values, and additional notes are given below the table. For free parameters, we give the starting value, minimum/maximum limits and $\Delta$, the step size used by \xspec to determine numerical derivatives during the fitting. We also show the parameters of the updated \optxconv SED model, which is substituted for \optxagnf in the discussion (Section \ref{subsub:spin}).}
	\small
	\centering
	\def\arraystretch{1.2}
	\begin{tabular}{ccclcccccc}
		\hline
		Model     & \# & Parameter    & Description                          & Free? & Start    & Min    & Max       & $\Delta$ & Units \\
		\hline
				
		\wabs     & 1  & \nh          & Galactic \Hi column density            & N & (A)        & --        & --         & --       & $10^{22} \; \rm cm^{-2}$ \\
		\\	
		\zwabs    & 1  & \nh          & Intrinsic \Hi column density           & Y & 0.0        & 0.0       & $10^5$     & 0.001    & $10^{22} \; \rm cm^{-2}$ \\
				  & 2  & $z$          & Redshift                               & N & (B)        & --        & --         & --       & -- \\
		\\		  
		\zdust    & 1  & Method       & Set reddening curve -- MW, SMC or LMC  & N & (C)        & --        & --         & --       & -- \\
		          & 2  & $E(B-V)$     & Intrinsic $B-V$ band extinction        & Y & 0.0        & 0.0       & 10.0       & 0.001    & mag \\
		          \vspace{-0.1cm}
	           	  & 3  & $R(V)$       & Ratio of absolute to selective         & N & (D)       & --        & --         & --       & -- \\
		          &    &              & extinction, $R(V)=A(V)/E(B-V)$         &  \\
		          & 4  & $z$          & Redshift                               & N & (B)       & --        & --         & --       & -- \\
		\\		
		\optxagnf & 1  & \M_BH        & BH mass                                & N$^{(a)}$ & (E)       & --        & --         & --       & \Msun \\
				  & 2  & $r_{\rm c}$  & Comoving distance                      & N & (F)       & --        & --         & --       & Mpc \\
				  & 3  & log(\mdot)   & Logarithm of Eddington fraction        & Y & 0.0        & $-$10.0   & 2.0        & 0.01     & -- \\
				  & 4  & $a_*$        & BH spin                                & N$^{(b)}$ & 0.0        & --        & --         & --       & -- \\
				  & 5  & \rcor        & Coronal radius                         & Y & 15.0       & 7.0       & 100.0      & 0.1      & \Rg \\
				  & 6  & log(\rout)   & Logarithm of outer AD radius           & Y$^{(c)}$ & 3.0        & 1.5       & 7.0        & 0.01     & log(\Rg) \\
				  & 7  & $kT_{\rm e}$ & SX electron temperature                & N & 0.2        & --        & --         & --       & keV \\
				  & 8  & $\tau$       & SX optical depth                       & N & 10         & --        & --         & --       & -- \\
				  & 9  & $\Gamma$     & PLT photon index                       & Y & 2.0        & 0.5       & 5.0        & 0.01     & -- \\
				  & 10 & $f_{\rm PLT}$ & Fraction of energy below \rcor in PLT (*) & Y & 0.3        & 0.0       & 1.0        & $10^{-6}$& -- \\
				  & 11 & $z$          & Redshift                               & N & (B)       & --        & --         & --       & -- \\
				  & 12 & Norm         & Arbitrary normalisation $= 1$          & N & 1.0        & --        & --         & --       & -- \\
		\\		
		\optxconv & 1--10             & \multicolumn{3}{c}{Same as \optxagnf}   \\
				  & 11 & $\theta$          & Inclination angle to observer          & N$^{(d)}$ & 60.0       & --        & --         & --       & deg.\ \\
				  & 12 & $z$          & Redshift                               & N & (B)       & --        & --         & --       & -- \\
				  & 13 & Norm         & Arbitrary normalisation $= 1$          & N & 1.0        & --        & --         & --       & -- \\

		\hline
		\multicolumn{9}{l}{(A) Value from \cite{kalberla05}.}   \\
		\multicolumn{9}{l}{(B) Measured in Paper I.}   \\
		\multicolumn{9}{l}{(C) Best fitting extinction curve used, as described in Section \ref{subsec:intred}.}   \\
		\multicolumn{9}{l}{(D) Fixed at 3.08, 3.16, 2.93 for MW, LMC, SMC respectively (\citealt{pei92}).}   \\
		\multicolumn{9}{l}{(E) Mass estimated from single epoch virial technique applied to \Ha (\citealt{greene05}), see Paper I.}   \\
		\multicolumn{9}{l}{(F) Calculated from $z$, assuming a flat cosmology with $H_0 = 70$ km s$^{-1}$ Mpc$^{-1}$, $\Omega_{\rm M} = 0.27$ and $\Omega_{\Lambda} = 0.73$.} \\
		\multicolumn{9}{l}{($a$) We test the impact of alterations from the mean \M_BH value in Section \ref{subsec:mass}.}   \\
		\multicolumn{9}{l}{($b$) Alternative spin values are explored in Section \ref{subsec:spin}.}   \\
		\multicolumn{9}{l}{($c$) The effect of fixing \rout to different values is tested in Section \ref{subsec:rout}.}   \\
		\multicolumn{9}{l}{($d$) Also tested $\theta = 0$, see Section \ref{subsub:spin}.}   \\
		\multicolumn{9}{l}{(*) Note that although \optxagnf requires $f_{\rm PLT}$, we quote $f_{\rm SX} = (1-f_{\rm PLT})$ throughout this work, for consistency with Paper I.} \\
			
	\end{tabular}
	\label{tab:modelpar}
\end{table*}

\label{lastpage}

\end{document}